\documentclass[12pt]{article}
\usepackage{epsfig}
\usepackage{amsmath}
\usepackage{cite}
\input amssym.def
\input amssym

\def\beq{\begin{equation}}
\def\eeq{\end{equation}}
\def\bea{\begin{eqnarray}}
\def\eea{\end{eqnarray}}

\def\eq#1{{Eq.~(\ref{#1})}}
\def\fig#1{{Fig.~\ref{#1}}}
\def\ud{\underline}

\newcommand{\as}{\alpha_s}
\newcommand{\am}{\alpha_\mu}

\newcommand{\abar}{{\bar \alpha}}
\newcommand{\oone}{
\begin{picture}(10,8)
\put(5,5){\circle{8}}
\put(2.9,2.5){{\scriptsize 1}}
\end{picture}
}

\newcommand{\un}[1]{\underline{#1}}

\begin{document}
\title{{\bf Solving The High Energy Evolution Equation \\[.5cm]
    Including Running Coupling Corrections \\[1.5cm] }} \author{ {\bf
    Javier L.  Albacete\thanks{e-mail:
      albacete@mps.ohio-state.edu}\hspace{0.2cm} and Yuri V.\ 
    Kovchegov\thanks{e-mail: yuri@mps.ohio-state.edu}}
  \\[.3cm] {\it\small Department of Physics, The Ohio State University}\\
  {\it\small Columbus, OH 43210,USA}\\[5mm]}

\date{April 2007}

\maketitle 

\thispagestyle{empty}

\begin{abstract}
  We study the solution of the nonlinear BK evolution equation with
  the recently calculated running coupling corrections
  \cite{Balitsky:2006wa,Kovchegov:2006vj}. Performing a numerical
  solution we confirm the earlier result of \cite{Albacete:2004gw}
  (obtained by exploring several possible scales for the running
  coupling) that the high energy evolution with the running coupling
  leads to a universal scaling behavior for the dipole-nucleus
  scattering amplitude, which is independent of the initial
  conditions. It is important to stress that the running coupling
  corrections calculated recently significantly change the shape of
  the scaling function as compared to the fixed coupling case, in
  particular leading to a considerable increase in the anomalous
  dimension and to a slow-down of the evolution with rapidity.  We
  then concentrate on elucidating the differences between the two
  recent calculations of the running coupling corrections. We explain
  that the difference is due to an extra contribution to the evolution
  kernel, referred to as the subtraction term, which arises when
  running coupling corrections are included. These subtraction terms
  were neglected in both recent calculations. We evaluate numerically
  the subtraction terms for both calculations, and demonstrate that
  when the subtraction terms are added back to the evolution kernels
  obtained in the two works the resulting dipole amplitudes agree with
  each other! We then use the complete running coupling kernel
  including the subtraction term to find the numerical solution of the
  resulting full non-linear evolution equation with the running
  coupling corrections. Again the scaling regime is recovered at very
  large rapidity with the scaling function unaltered by the
  subtraction term.
\end{abstract}

\thispagestyle{empty}

\newpage

\setcounter{page}{1}

%%%%%%%%%%%%%%%%%%%%%%%%%%%%%%%%%%%%%%%%%%%%%%%%%%%%%%%%%%%%%%%%%%%%%%%%%%%%%

\section{Introduction} \label{intro}

Recently our understanding of the linear
Balitsky--Fadin--Kuraev--Lipatov (BFKL) \cite{Kuraev:1977fs,Bal-Lip}
and non-linear
Jalilian-Marian--Iancu--McLerran--Weigert--Leonidov--Kovner (JIMWLK)
\cite{Jalilian-Marian:1997jx, Jalilian-Marian:1997gr,
  Jalilian-Marian:1997dw, Jalilian-Marian:1998cb, Kovner:2000pt,
  Weigert:2000gi, Iancu:2000hn,Ferreiro:2001qy} and Balitsky-Kovchegov
(BK) \cite{Balitsky:1996ub,
  Balitsky:1997mk,Balitsky:1998ya,Kovchegov:1999yj, Kovchegov:1999ua}
small-$x$ evolution equations in the Color Glass Condensate
\cite{Gribov:1981ac,Mueller:1986wy,McLerran:1994vd,McLerran:1993ka,
  McLerran:1993ni,Kovchegov:1996ty,Kovchegov:1997pc,Jalilian-Marian:1997xn,
  Jalilian-Marian:1997jx, Jalilian-Marian:1997gr,
  Jalilian-Marian:1997dw, Jalilian-Marian:1998cb, Kovner:2000pt,
  Weigert:2000gi, Iancu:2000hn,Ferreiro:2001qy,Kovchegov:1999yj,
  Kovchegov:1999ua, Balitsky:1996ub, Balitsky:1997mk,
  Balitsky:1998ya,Iancu:2003xm,Weigert:2005us,Jalilian-Marian:2005jf}
has been improved due to the completion of the calculations
determining the scale of the running coupling in the evolution kernel
in
\cite{Balitsky:2006wa,Gardi:2006rp,Kovchegov:2006vj,Kovchegov:2006wf}.
The calculations in \cite{Balitsky:2006wa,Kovchegov:2006vj} proceeded
by including $\as \, N_f$ corrections into the evolution kernel and by
then completing $N_f$ to the complete one-loop QCD beta-function by
replacing $N_f \rightarrow - 6 \pi \beta_2$. Calculation of the $\as
\, N_f$ corrections is particularly easy in the $s$-channel light-cone
perturbation theory formalism
\cite{Lepage:1980fj,Brodsky:1997de} used to derive the BK and JIMWLK
equations: there $\as \, N_f$ corrections are solely due to chains of
quark bubbles placed onto the $s$-channel gluon lines.

The analytical results of \cite{Balitsky:2006wa,Kovchegov:2006vj} are
not very concise and could not have been guessed without an explicit
calculation.  After finding $\as \, N_f$ corrections, the obtained
contributions had to be divided into the running coupling part, which
has a form of a running coupling correction to the leading-order (LO)
JIMWLK or BK kernel, and into the "subtraction piece", which would
bring in new structures into the kernel. Such separation had to be
done both in \cite{Balitsky:2006wa} and in \cite{Kovchegov:2006vj}.
Unfortunately, there appears to be no unique way to perform this
separation: it is not surprising, therefore, that it was done
differently in both papers \cite{Balitsky:2006wa,Kovchegov:2006vj}.
This resulted in two different running coupling terms, shown below in
Eqs. (\ref{bal_run}) and (\ref{kw_run}) along with Eqs. (\ref{kbal})
and (\ref{kkw}). Such a discrepancy has led to a misconception in the
community that the calculations of \cite{Balitsky:2006wa} and of
\cite{Kovchegov:2006vj} disagree at some fundamental level.

Indeed to compare the results of \cite{Balitsky:2006wa} and
\cite{Kovchegov:2006vj} one has to undo the separation into the
running coupling and subtraction terms: combining both terms one
should compare full kernels of the evolution equation obtained in
\cite{Balitsky:2006wa,Kovchegov:2006vj}. There is another more
physical reason to perform such comparison: in principle, there is no
small parameter making the subtraction term smaller than the running
coupling term and thus justifying neglecting the former compared to
the latter. Even the labeling of one term as ``running coupling''
piece is somewhat misleading, since it may give an impression that the
neglected subtraction term has no running coupling corrections in it.
As was shown in \cite{Kovchegov:2006wf} both terms actually contribute
to the running coupling corrections to the BFKL equation (if one uses
the separation of \cite{Kovchegov:2006vj} to define the terms).

In this paper we perform numerical analysis of the BK evolution
equation with the $\as \, N_f$ corrections resummed to all orders and
with $N_f$ completed to the QCD beta-function, $N_f \rightarrow - 6
\pi \beta_2$, with $\beta_2$ given in \eq{beta}. We first solve the BK
equation keeping the running coupling term only, with the kernels
given by Eqs.  (\ref{kbal}) and (\ref{kkw}). Indeed the solutions we
find this way are different from each other. We then evaluate the
subtraction terms for both cases and show that inclusion of
subtraction terms puts the results of \cite{Balitsky:2006wa} and
\cite{Kovchegov:2006vj} in perfect agreement with each other! We
complete our analysis by solving the BK equation with the full kernel
including both the running coupling and subtraction terms.

This work is structured as follows. Section \ref{secsub} begins with
Sect. \ref{gen_conc} in which we review the $\as \, N_f$ corrections
to the dipole scattering amplitude evolution equation recently derived
in \cite{Balitsky:2006wa,Kovchegov:2006vj} and the subtraction method
employed in both works to separate the running coupling contributions
from the subtraction terms. We discuss the scheme dependence of the
running coupling terms introduced by this separation. We proceed in
Sect. \ref{derivation} by deriving the explicit expressions for the
subtracted terms. The calculation is based on the results of
\cite{Kovchegov:2006vj}. Our analytical results are summarized in
Sect. \ref{analyt_sum}, where we give the explicit final expression
for the kernel of the subtraction term in \eq{K1run_final}, which,
combined with \eq{fullBK} gives us the subtraction terms
(\ref{sub_full_bal}) and (\ref{sub_full_kw}) for the subtractions
performed in \cite{Balitsky:2006wa} and in \cite{Kovchegov:2006vj}
correspondingly.

In Sect. \ref{setup} we explain the numerical method we use to solve
the evolution equations. We also list the initial conditions used,
along with the definition of the saturation scale employed. Throughout
the paper we will avoid the important question of the Landau pole and
the contribution of renormalons to small-$x$ evolution. As we explain
in Sect. \ref{setup}, we will simply ``freeze'' the running coupling
at a constant value in the infrared. For a detailed study of the
renormalon effects in the non-linear evolution we refer the readers to
\cite{Gardi:2006rp}.

Our numerical results are presented in Sect. \ref{results}.  By
solving the evolution equations with the running coupling term only in
Sect.  \ref{running} we show that the resulting dipole amplitude
differs significantly from the fixed coupling case. We also observe
that the amplitude obtained by solving the equation obtained in
\cite{Kovchegov:2006vj} is very close to the result of solving the BK
evolution with a postulated parent-dipole running of the coupling
constant. Both these amplitudes are quite different from the solution
of the equation derived in \cite{Balitsky:2006wa}, as one can see from
Fig. \ref{sols}. In spite of that, all three evolution equations
studied (the ones derived in \cite{Balitsky:2006wa},
\cite{Kovchegov:2006vj} and the parent-dipole running coupling model)
give approximately identical scaling function for the dipole amplitude
at high rapidity, as demonstrated in \fig{scal} in Sect.
\ref{scaling}. It is worth noting that, as can be seen from
\fig{anom}, the anomalous dimension we extracted from our solution is
$\gamma \approx 0.85$, which is different from the fixed coupling
anomalous dimension of $\gamma \approx 0.64$. The former anomalous
dimension also appears to disagree with the predictions of analytical
approximations to the behavior of the dipole amplitude with running
coupling from
\cite{Mueller:2002zm,Triantafyllopoulos:2002nz,Iancu:2002tr,Munier:2004xu,Beuf:2007cw}.
In Sect. \ref{subterm} we numerically evaluate the subtraction terms
for both \cite{Balitsky:2006wa} and \cite{Kovchegov:2006vj} and show
that their contributions are important, as shown in \fig{sub1}.
However, subtraction terms decrease with increasing rapidity, such
that at high enough rapidities their relative contribution becomes
small (see \fig{subr}). In \fig{subt} we show that inclusion of
subtraction terms makes the results of \cite{Balitsky:2006wa} and
\cite{Kovchegov:2006vj} agree with each other. Finally, the numerical
solution of the full (all orders in $\as \, \beta_2$) evolution
equation including both the running coupling and subtraction terms is
performed in Sect. \ref{comprc}. The results are shown in \fig{subev}.
All the main features of the evolution with the running coupling are
preserved in the full solution: the growth of the dipole amplitude and
of the saturation scale with rapidity is slowed down (for the latter
see \fig{subqs}).  The scaling function of \fig{scal} is unaltered by
the subtraction term, as shown in \fig{scalsub}. 

We summarize and discuss our main conclusions in Sect.
\ref{conclusions}.
 
%%%%%%%%%%%%%%%%%%%%%%%%%%%%%%%%%%%%%%%%%%%%%%%%%%%%%%%%%%%%%%%%%%%%%%%%%%%%%

\section{Scheme dependence} \label{secsub}

\subsection{Inclusion of running coupling corrections: general concepts}
\label{gen_conc}

The BK evolution equation for the dipole scattering matrix reads
\begin{align}
  \frac{\partial S(\ud{x}_0,\ud{x}_1;Y)}{\partial Y}=
\int d^2 z \,K(\ud{x}_0,\ud{x}_1,\ud{z})
\left[S(\ud{x}_0,\ud{z};Y)\,S(\ud{z},\ud{x}_1;Y)-S(\ud{x}_0,\ud{x}_1;Y)\right]\,,       
\label{llbk}
\end{align}
where 
\begin{align}
  K(\ud{x}_0,\ud{x}_1,\ud{z})=\frac{\as \, N_c}{2\pi^2}\frac{r^2}{r_1^2\,r^2_2}
\end{align}
is the kernel of the evolution. Here transverse two-dimensional
vectors ${\un x}_0$ and ${\un x}_1$ denote the transverse coordinates
of the quark and the anti-quark in the parent dipole, while $\un z$ is
the position of the gluon produced in one step of evolution
\cite{Mueller:1994rr,Mueller:1994jq,Mueller:1995gb,Chen:1995pa}.  We
have introduced the notation $\ud{r}=\ud{x}_0-\ud{x}_1$,
$\ud{r}_1=\ud{x}_0-\ud{z}$, $\ud{r}_2=\ud{z}-\ud{x}_1$ for the sizes
of the parent and of the new (daughter) dipoles created by one step of
the evolution. The notation $r \equiv |\ud{r}|$ for all the
2-dimensional vectors will be also employed throughout the rest of the
paper. \eq{llbk} admits a clear physical interpretation: the original
parent dipole, when boosted to higher rapidities, may emit a new gluon
which, in the large-$N_c$ limit, is equivalent to a quark-antiquark
pair.  Thus, the original dipole splits into two new dipoles sharing a
common transverse coordinate: the transverse position of the emitted
gluon, $\ud{z}$.  The nonlinear term in the right hand side of
\eq{llbk} accounts for either one of the two new dipoles interacting
with the target, along with the possibilities of only one dipole
interacting or no interaction at all, while the subtracted linear term
reflects virtual corrections. The kernel of the evolution is just the
probability of one gluon emission calculated at leading logarithmic
accuracy in $\alpha_s\ln(1/x_B)$, where $x_B$ is the fraction of
momentum carried by the emitted gluon
\cite{Mueller:1994rr,Mueller:1994jq,Mueller:1995gb,Chen:1995pa}.
 
Under the eikonal approximation the dipole scattering matrix off a
hadronic target at a fixed rapidity is given by the average over the
hadron field configurations of Wilson lines $V$ calculated along fixed
transverse coordinates (those of the quark and of the antiquark). More
specifically 
\beq 
S(\ud{x}_0,\ud{x}_1;Y)=\frac{1}{N_c}\langle
\mbox{tr} \left\{V(\ud{x}_0)V^{\dagger}(\ud{x}_1)\right\}\rangle\,.
\label{sdip}
\eeq
Hence, the integrand of \eq{llbk} can be regarded as a three point 
function in the sense that the gluon fields 
of the target are evaluated at three different transverse positions,
those of the original quark and antiquark plus the one of the emitted
gluon. 

\begin{figure}[ht]
\begin{center}
\includegraphics[height=4cm]{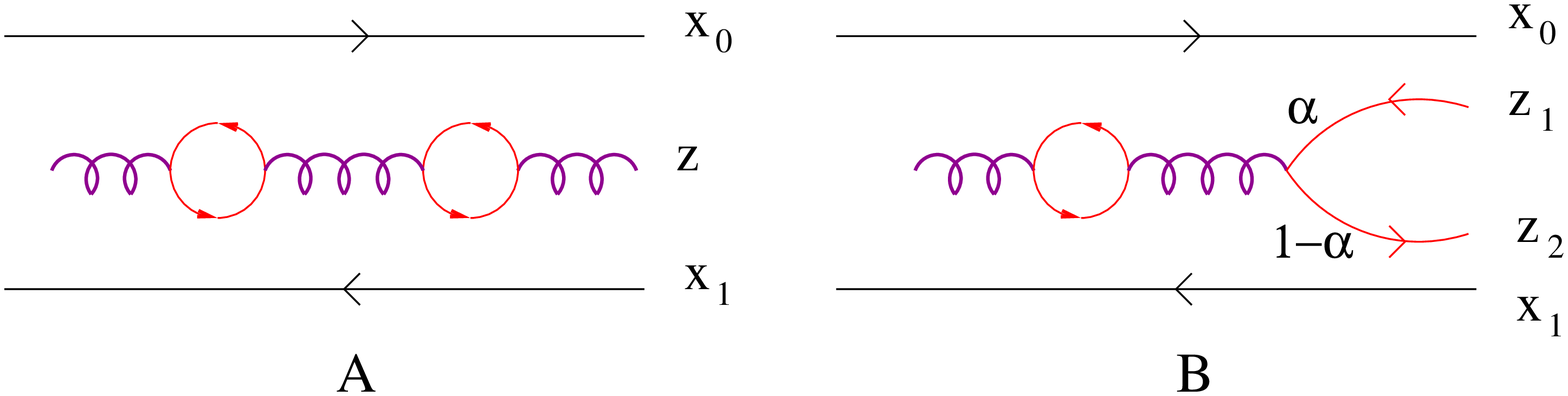}%,width=14.5cm]{c1.eps}
\end{center}
\caption{Schematic representation of the diagrams contributing to
  quark-NLO evolution.}
\label{nlodip}
\end{figure}

However, the inclusion of higher order corrections to the evolution
equation via all order resummation of $\alpha_s \,N_f$ contributions
as recently derived in \cite{Balitsky:2006wa,Kovchegov:2006vj} brings
in new physical channels that modify the three point structure of the
leading-log equation. The dipole structure generated under evolution
by diagrams like the one depicted in \fig{nlodip}A (for more detailed
discussion of the diagrammatic content of the high order corrections
see \cite{Kovchegov:2006vj}) is identical to the one previously
discussed for the leading order equation, the only novelty being that
the propagator of the emitted gluon is now dressed with quark loops,
modifying the emission probability but leaving untouched the
interaction terms. On the contrary, diagrams like the one in
\fig{nlodip}B in which a quark-antiquark pair (rather than a gluon) is
added to the evolved wave function modify the interaction structure of
the evolution equation. The evolution of the parent dipole scattering
matrix driven by these kind of terms is proportional to the scattering
matrix of the two newly created dipoles (the one formed by the
original quark and the new antiquark and vice versa), $\sim
S(\ud{x}_0,\ud{z}_1)S(\ud{z}_2,\ud{x}_1)$. This term depends on four
different transverse coordinates, i.e., it is a four point function
and, therefore, its contribution to the evolution equation cannot be
accounted for by a mere modification of the emission kernel of the
leading order equation.
 
%%%%%%%%%%%%%%%%%%%%%%%%%%%%%
%%%%%%%%%%%%%%%%%%%%%%%%%%%%%
To discuss in more detail the modifications introduced by the high
order corrections, we find it useful to rewrite the evolution equation
in the following, rather general way 
\begin{align}
  \frac{\partial  S(\ud{x}_0,\ud{x}_1;Y)}{\partial Y}=
\mathcal{F}\left[S(\ud{x}_0,\ud{x}_1;Y)\right]
\label{formal}
\end{align}
where $\mathcal{F}$ is a functional of the dipole scattering
matrix which for the original derivation of the equation is given by
the right hand side of \eq{llbk}.  In general it can be
decomposed into two pieces 
\begin{align}
  \mathcal{F}\left[S\right]=\mathcal{R}\left[S\right]-\mathcal{S}\left[S\right]\,.
\label{f}
\end{align}

The first term, $\mathcal{R}$, which we will call the 'running
coupling' contribution, gathers all the higher order in $\as \, N_f$
corrections to the evolution that can be recast in a functional form
that looks identical to the leading order one but with a modified
kernel, $\tilde{K}$, which includes all the terms setting the scale
for the running coupling:
\begin{align}
  \mathcal{R}\left[S(\ud{x}_0,\ud{x}_1;Y)\right]= \int\,d^2 z
  \,\tilde{K}(\ud{x}_0,\ud{x}_1, {\un z})
  \left[S(\ud{x}_0,\ud{z};Y)\,S(\ud{z},\ud{x}_1;Y)-S(\ud{x}_0,\ud{x}_1;Y)\right].
\label{run}
\end{align}

The second term, $\mathcal{S}$, henceforth referred to as the
'subtraction' contribution, encodes those contributions that depart
from the three point structure of the leading-log equation. The
explicit derivation and expressions for this term are presented in the
next section. The relative minus sign between the two terms in \eq{f}
has been introduced for latter convenience.

%%%%%%%%%%%%%%%%%%%%%%%%%%%%%
\begin{figure}[ht]
\begin{center}
\includegraphics[height=8.5cm]{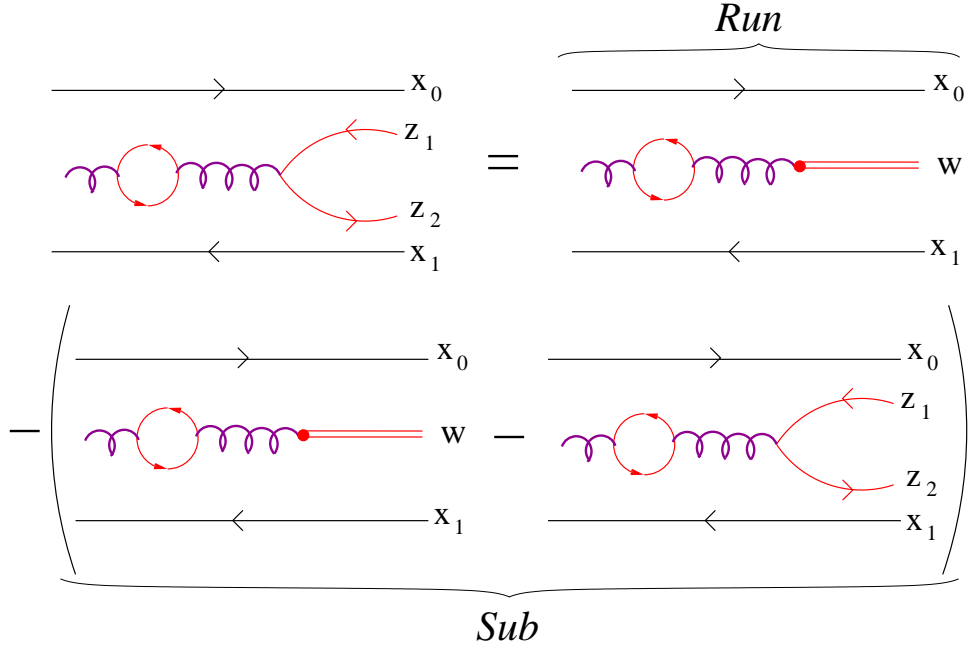}%,width=14.5cm]{c1.eps}
\end{center}
\caption{Schematic representation of the subtraction procedure.}
\label{scheme}
\end{figure}
%%%%%%%%%%%%%%%%%%%%%%%%%%%%%

Importantly, the decomposition of $\mathcal{F}$ into running coupling
and subtraction contributions, although constrained by unitarity
arguments, is not unique.  Two different separation schemes have been
proposed in \cite{Kovchegov:2006vj,Balitsky:2006wa}.  They are both
based on a similar strategy, sketched in \fig{scheme}, that can be
summarized as follows. The newly created quark-antiquark pair added to
the wave function in the diagrams \fig{nlodip}B is shrunk to a point,
called the subtraction point, by integrating out one of the
coordinates in the dipole-$q\bar q$ wave function, rendering the
previously discussed four point nature of these contributions into a
three point one. This integrated three point contribution is added to
the running coupling contribution, whereas the original four point
term minus its integrated version are assigned to the subtraction
contribution. The divergence between the two approaches stems from the
choice of the subtraction point.  In the subtraction scheme proposed
by Balitsky in \cite{Balitsky:2006wa} the subtraction point is chosen
to be the transverse coordinate of either the quark, ${\un z}_2$, or
the antiquark, ${\un z}_1$. The kernel for the running coupling
functional, \eq{run}, obtained in this way is
\begin{align}
  \tilde{K}^{\text{Bal}}(\ud{r},\ud{r}_1,\ud{r}_2)=\frac{N_c\,\alpha_s(r^2)}{2\pi^2}
  \left[\frac{r^2}{r_1^2\,r_2^2}+
    \frac{1}{r_1^2}\left(\frac{\alpha_s(r_1^2)}{\alpha_s(r_2^2)}-1\right)+
    \frac{1}{r_2^2}\left(\frac{\alpha_s(r_2^2)}{\alpha_s(r_1^2)}-1\right)
  \right]\,.
\label{kbal}
\end{align}
On the other hand, in the subtraction procedure followed in
\cite{Kovchegov:2006vj} (which we will refer to as KW) the zero size
quark-antiquark pair is fixed at the transverse coordinate of the
gluon, $z=\alpha z_1+(1-\alpha)z_2$, where $\alpha$ is the fraction of
the gluon's longitudinal momentum carried by the quark, yielding the
following expression for the kernel of the running coupling
contribution:
\begin{align}
  \tilde{K}^{\text{KW}}(\ud{r},\ud{r}_1,\ud{r}_2)=\frac{N_c}{2\pi^2}\left[
    \alpha_s(r_1^2)\frac{1}{r_1^2}-
    2\,\frac{\alpha_s(r_1^2)\,\alpha_s(r_2^2)}{\alpha_s(R^2)}\,\frac{
      \ud{r}_1\cdot \ud{r}_2}{r_1^2\,r_2^2}+
    \alpha_s(r_2^2)\frac{1}{r_2^2} \right]\,,
\label{kkw}
\end{align}
where
\beq
R^2(\ud{r},\ud{r}_1,\ud{r}_2)=r_1\,r_2\left(\frac{r_2}{r_1}\right)^
{\frac{r_1^2+r_2^2}{r_1^2-r_2^2}-2\,\frac{r_1^2\,r_2^2}{
      \ud{r}_1\cdot\ud{r}_2}\frac{1}{r_1^2-r_2^2}}\,.  
\label{r}
\eeq 
As we shall discuss later, the scheme dependence originated by
the choice of the subtraction point is substantial and has an
important effect in the solutions of the evolution equation when only
the running contribution is taken into account.

In our numerical study we will also consider the following ad hoc
prescription for the kernel of the running coupling functional in
which the scale for the running of the coupling is set to be the size
of the parent dipole \beq
\tilde{K}^{pd}(\ud{r},\ud{r}_1,\ud{r}_2)=\frac{N_c\,
  \alpha_s(r^2)}{2\pi^2}\frac{r^2}{r_1^2\,r_2^2}.
\label{kpd}
\eeq 
This prescription is useful as a benchmark used to compare with
previous numerical \cite{Albacete:2004gw} and analytical works
\cite{Munier:2003sj,Mueller:2002zm,Triantafyllopoulos:2002nz} where
this ansatz was used.

%%%%%%%%%%%%%%%%%%%%%%%%%%%%%%%%%%%%%%%%%%%%%%%%%%%%%%%%%%%%%%%%%%%%%%%%%%%%%%%%%

\subsection{Derivation of the subtraction term}
\label{derivation}

\begin{figure}[ht]
\begin{center}
\includegraphics[width=15cm]{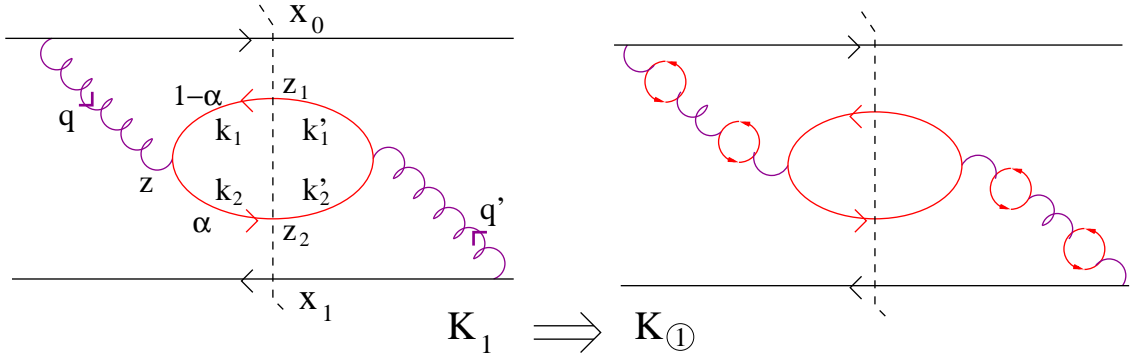}
\end{center}
\caption{A lowest order leading-$N_f$ NLO correction which gives rise 
  to the subtraction term is shown on the left. The same diagram with
  the gluon lines ``dressed'' by chains of fermion bubbles, as shown
  on the right, gives the full (resumming all powers of $\am \, N_f$)
  contribution to the subtraction term. Calculation of the subtraction
  term is pictured in \fig{scheme}.}
\label{kersub}
\end{figure}

We begin by considering the NLO contribution to the kernel of the
JIMWLK and BK evolution equations with the $s$-channel gluon splitting
into a quark-antiquark pair, which then interacts with the target, as
shown on the left hand side of \fig{kersub}. The contribution of this
diagram has been calculated in \cite{Kovchegov:2006vj}. The resulting
JIMWLK kernel is \cite{Kovchegov:2006vj}
\begin{align}\label{K1init}
  {\cal K}_1^{\text{NLO}} ({\ud x}_0, {\ud x}_1 ; {\ud z}_1, {\ud z}_2) \,
  = & \, 4 \, %\am^2 \, 
  N_f \, \int\limits_0^1 d \alpha \, \int \frac{d^2
    k}{(2\pi)^2}\frac{d^2 k'}{(2\pi)^2}\frac{d^2 q}{(2\pi)^2}
  \frac{d^2 q'}{(2\pi)^2} \ e^{ -i {\ud q}\cdot ({\ud z}-{\ud x}_0) +i
    {\ud q}' \cdot ({\ud z}-{\ud x}_1) -i({\ud k}-{\ud k}') \cdot {\ud
      z}_{12}} \notag \\ & \times \left[ \frac{1}{{\ud q}^2{\ud
        q}'^{2}} \frac{(1-2\alpha)^2 {\ud q}\cdot{\ud k}\ {\ud k}'
      \cdot {\ud q}' + {\ud q} \cdot {\ud q}' \ {\ud k}\cdot{\ud k}' -
      {\ud q}\cdot{\ud k}' \ {\ud k}\cdot{\ud q}'}{\Big[{\ud k}^2+{\ud
        q}^2\alpha(1-\alpha)\Big]\Big[{\ud k}'^2+{\ud
        q}'^2\alpha(1-\alpha)\Big]} \right. \notag \\ & + \frac{2 \,
    \alpha \, (1-\alpha) \, (1-2 \alpha)}{\Big[{\ud k}^2+{\ud
      q}^2\alpha(1-\alpha)\Big]\Big[{\ud k}'^2+{\ud
      q}'^2\alpha(1-\alpha)\Big]} \, \left( \frac{{\ud k} \cdot {\ud
        q}}{{\ud q}^2} + \frac{{\ud k}' \cdot {\ud q}'}{{\ud q}'^2}
  \right)\notag \\ & \left. + \, \frac{4 \, \alpha^2 \,
      (1-\alpha)^2}{\Big[{\ud k}^2+{\ud
        q}^2\alpha(1-\alpha)\Big]\Big[{\ud k}'^2+{\ud
        q}'^2\alpha(1-\alpha)\Big]} \right].
\end{align}
The momentum labels in the above equation are explained on the left
hand side of \fig{kersub}. If ${\un k}_1$ and ${\un k}_2$ are the
transverse momenta of the quark and of the antiquark in the produced
pair as shown in \fig{kersub}, then the transverse momentum of the
gluon is ${\un q} = {\un k}_1 + {\un k}_2$. The other transverse
momentum we use is ${\un k}={\un k}_1 (1-\alpha)-{\un k}_2\alpha$,
where $\alpha$ is the fraction the of gluon's ``plus'' momentum
carried by the quark, $\alpha \equiv k_{1+} / (k_{1+} + k_{2+})$. The
prime over the transverse momentum denotes the momentum of the same
particle in the complex conjugate amplitude.  For instance ${\un q}'$
is the momentum of the $s$-channel gluon in the complex conjugate
amplitude. Finally, ${\un z}_1$ and ${\un z}_2$ denote the transverse
coordinates of the quark and the antiquark. In \eq{K1init} we use
${\un z}_{12} = {\un z}_1 - {\un z}_2$ (the transverse separation
between the quark and the antiquark) and ${\un z} = \alpha \, {\un
  z}_1 + (1-\alpha) \, {\un z}_2$ (the transverse coordinate of the
gluon).

%%%%%%%%%%%%%%%%%%%%%%%%%%%%%
%%%%%%%%%%%%%%%%%%%%%%%%%%%%%

To obtain the BK kernel from \eq{K1init} one should sum over all
possible emissions of the gluon off the quark and antiquark lines in
the incoming dipole both in the amplitude and in the complex conjugate
amplitude, which is accomplished by
\begin{align}\label{K1BK}
  K_1^{\text{NLO}} ({\un x}_0, {\un x}_1 ; {\un z}_1, {\un z}_2) \, = \,
  C_F \, \sum_{m,n = 0}^1 \, (-1)^{m+n} \, {\cal K}_1^{\text{NLO}}
  & ({\un x}_m, {\un x}_n ; {\un z}_1, {\un z}_2).
\end{align}
Below we will label the JIMWLK kernel by calligraphic letter ${\cal K}$
and the corresponding BK kernel by $K$.

The contribution of the kernel from \eq{K1BK} to the right hand side
of the NLO version of \eq{llbk} is given by the following term
\begin{align}\label{subtr1}
  \am^2 \, \int d^2 z_1 \, d^2 z_2 \, K_1^{\text{NLO}} ({\un x}_0,
  {\un x}_1 ; {\un z}_1, {\un z}_2) \, S ({\un x}_{0}, {\un z}_1, Y)
  \, S ({\un z}_{2}, {\un x}_1, Y)
\end{align}
with $\am$ the bare coupling.

As shown in \fig{scheme}, at the NLO level, the subtraction term
introduced in \eq{f}, is then defined by
\begin{align}\label{sub_def}
  {\mathcal S}_{\text{NLO}} [S] \, = \, \am^2 \, & \int d^2 z_1 \, d^2
  z_2 \, K_1^{\text{NLO}} ({\un x}_0, {\un x}_1 ; {\un z}_1, {\un
    z}_2) \notag \\ \times \, & [ S ({\un x}_{0}, {\un w}, Y) \, S
  ({\un w}, {\un x}_1, Y) - S ({\un x}_{0}, {\un z}_1, Y) \, S ({\un
    z}_{2}, {\un x}_1, Y)],
\end{align}
where $\un w$ is the point of subtraction in the transverse coordinate
space. In \cite{Balitsky:2006wa} it was chosen to be equal to the
transverse coordinate of either the quark or the antiquark,
\begin{align}
  {\un w} = {\un z}_1 \hspace*{1cm} \mbox{or} \hspace*{1cm} {\un w} = {\un z}_2,
\end{align}
as both choices lead to the same subtraction term ${\mathcal
  S}^{\text{Bal}}_{\text{NLO}} [S]$:
\begin{align}
  \mathcal{S}^{\text{Bal}}_{\text{NLO}} [S] \, = \, & \int d^2
  z_1\,d^2 z_2\,K_1^{\text{NLO}} (\ud{x}_0,\ud{x}_1;\ud{z}_1,\ud{z}_2)
  \notag \\ \times \, & \left[ S(\ud{x}_0,\ud{z}_1, Y) \,
    S(\ud{z}_1,\ud{x}_1, Y) - S(\ud{x}_0,\ud{z}_1, Y) \,
    S(\ud{z}_2,\ud{x}_1, Y) \right].
\label{subkw}
\end{align}

In \cite{Kovchegov:2006vj} the subtraction point was chosen to be the
transverse coordinate of the gluon $\un z$,
\begin{align}
  {\un w} = {\un z} = \alpha \, {\un z}_1 + (1-\alpha) \, {\un z}_2.
\end{align}
This leads to the following subtraction term, which we denote ${\mathcal
  S}^{\text{KW}}_{\text{NLO}} [S]$:
\begin{align}
  \mathcal{S}^{\text{KW}}_{\text{NLO}} [S]\, = \, & \int d^2 z_1 \,
  d^2 z_2\,K_1^{\text{NLO}} (\ud{x}_0,\ud{x}_1;\ud{z}_1,\ud{z}_2)
  \notag \\ \times \, & \left[\,S(\ud{x}_0,\ud{z}, Y) \,
    S(\ud{z},\ud{x}_1, Y) - S(\ud{x}_0,\ud{z}_1, Y) \,
    S(\ud{z}_2,\ud{x}_1, Y) \right].
\label{subbal}
\end{align}

Indeed the complete kernel in \eq{f} is independent of the choice of
$\un w$.  However, since the subtraction term of \eq{sub_def} was
neglected both in \cite{Balitsky:2006wa} and in
\cite{Kovchegov:2006vj}, different choices of $\un w$ led to different
expressions for the remaining running coupling part ${\cal R} [S]$,
i.e., to different answers as far as investigations in
\cite{Balitsky:2006wa} and in \cite{Kovchegov:2006vj} were concerned.
Different choice of $\un w$ is the main source of the discrepancy of
final answers of \cite{Balitsky:2006wa} and \cite{Kovchegov:2006vj},
though it does not imply any disagreement in the full expression
(\ref{f}).

Our goal in this Section is to evaluate $K_1^{\text{NLO}}
(\ud{x}_0,\ud{x}_1;\ud{z}_1,\ud{z}_2)$ from \eq{K1init} including the
running coupling corrections. The $s$-channel light-cone perturbation
theory formalism makes such inclusion simple \cite{Kovchegov:2006vj}:
all we have to do is include infinite chains of quark bubbles on the
gluon lines in the amplitude and in the complex conjugate amplitude,
as depicted on the right hand side of \fig{kersub}. Performing
calculations similar to those done in \cite{Kovchegov:2006vj} one
arrives at
\begin{align}\label{K1run}
  {\cal K}_{\oone} ({\ud x}_0, {\ud x}_1 ; {\ud z}_1, {\ud z}_2) \, =
  & \, 4 \, N_f \, \int\limits_0^1 d \alpha \, \int \frac{d^2
    k}{(2\pi)^2}\frac{d^2 k'}{(2\pi)^2}\frac{d^2 q}{(2\pi)^2}
  \frac{d^2 q'}{(2\pi)^2} \ e^{ -i {\ud q}\cdot ({\ud z}-{\ud x}_0) +i
    {\ud q}' \cdot ({\ud z}-{\ud x}_1) -i({\ud k}-{\ud k}') \cdot {\ud
      z}_{12}} \notag \\ & \times \left[ \frac{1}{{\ud q}^2{\ud
        q}'^{2}} \frac{(1-2\alpha)^2 {\ud q}\cdot{\ud k}\ {\ud k}'
      \cdot {\ud q}' + {\ud q} \cdot {\ud q}' \ {\ud k}\cdot{\ud k}' -
      {\ud q}\cdot{\ud k}' \ {\ud k}\cdot{\ud q}'}{\Big[{\ud k}^2+{\ud
        q}^2\alpha(1-\alpha)\Big]\Big[{\ud k}'^2+{\ud
        q}'^2\alpha(1-\alpha)\Big]} \right. \notag \\ & + \frac{2 \,
    \alpha \, (1-\alpha) \, (1-2 \alpha)}{\Big[{\ud k}^2+{\ud
      q}^2\alpha(1-\alpha)\Big]\Big[{\ud k}'^2+{\ud
      q}'^2\alpha(1-\alpha)\Big]} \, \left( \frac{{\ud k} \cdot {\ud
        q}}{{\ud q}^2} + \frac{{\ud k}' \cdot {\ud q}'}{{\ud q}'^2}
  \right)\notag \\ & \left. + \, \frac{4 \, \alpha^2 \,
      (1-\alpha)^2}{\Big[{\ud k}^2+{\ud
        q}^2\alpha(1-\alpha)\Big]\Big[{\ud k}'^2+{\ud
        q}'^2\alpha(1-\alpha)\Big]} \right] \notag \\ & \times
  \frac{1}{\left( 1 + \am \beta_2 \ln \frac{{\un q}^2 \,
        e^{-5/3}}{\mu_{\overline{{\text{MS}}}}^2} \right) \, \left( 1
      + \am \beta_2 \ln \frac{{\un q}'^2 \,
        e^{-5/3}}{\mu_{\overline{{\text{MS}}}}^2} \right)}
\end{align}
where ${\cal K}_{\oone}$ denotes the kernel with the running coupling
corrections resummed to all orders. Just like in
\cite{Kovchegov:2006vj,Kovchegov:2006wf}, here we will use
$\overline{\text{MS}}$ renormalization scheme. Inclusion of fermion
bubble chains generated two denominators at the end of \eq{K1run},
which is its only difference from \eq{K1init}. Here
\begin{align}\label{beta}
  \beta_2 = \frac{11 N_c - 2 N_f}{12 \, \pi}. 
\end{align}

Now we have to perform the transverse momentum integrals in
\eq{K1run}. First we expand the denominators at the end of \eq{K1run}
into a power series and rewrite \eq{K1run} as
\begin{align}\label{K1run1}
  {\cal K}_{\oone} ({\ud x}_0, {\ud x}_1 ; {\ud z}_1, {\ud z}_2) \, =
  & \, 4 \, N_f \, \sum\limits_{n,m =0}^\infty (- \am \beta_2)^{n+m}
  \, \frac{d^n}{d \lambda^n} \, \frac{d^m}{d \lambda^{\prime m}} \,
  \left\{ \, \int\limits_0^1 d \alpha \, \int \frac{d^2
      k}{(2\pi)^2}\frac{d^2 k'}{(2\pi)^2}\frac{d^2 q}{(2\pi)^2}
    \frac{d^2 q'}{(2\pi)^2} \right. \notag \\ & \, e^{ -i {\ud q}\cdot
    ({\ud z}-{\ud x}_0) +i {\ud q}' \cdot ({\ud z}-{\ud x}_1) -i({\ud
      k}-{\ud k}') \cdot {\ud z}_{12}} \, \left( \frac{q^2}{\mu^2}
  \right)^\lambda \, \left(
    \frac{q'^2}{\mu^2} \right)^{\lambda'} \notag \\
  & \times \left[ \frac{1}{{\ud q}^2{\ud q}'^{2}} \frac{(1-2\alpha)^2
      {\ud q}\cdot{\ud k}\ {\ud k}' \cdot {\ud q}' + {\ud q} \cdot
      {\ud q}' \ {\ud k}\cdot{\ud k}' - {\ud q}\cdot{\ud k}' \ {\ud
        k}\cdot{\ud q}'}{\Big[{\ud k}^2+{\ud
        q}^2\alpha(1-\alpha)\Big]\Big[{\ud k}'^2+{\ud
        q}'^2\alpha(1-\alpha)\Big]} \right. \notag \\ \displaybreak[0]
  & + \frac{2 \, \alpha \, (1-\alpha) \, (1-2 \alpha)}{\Big[{\ud
      k}^2+{\ud q}^2\alpha(1-\alpha)\Big]\Big[{\ud k}'^2+{\ud
      q}'^2\alpha(1-\alpha)\Big]} \, \left( \frac{{\ud k} \cdot {\ud
        q}}{{\ud q}^2} + \frac{{\ud k}' \cdot {\ud q}'}{{\ud q}'^2}
  \right) \notag \\ & + \left. \left.  \left. \frac{4 \, \alpha^2 \,
          (1-\alpha)^2}{\Big[{\ud k}^2+{\ud
            q}^2\alpha(1-\alpha)\Big]\Big[{\ud k}'^2+{\ud
            q}'^2\alpha(1-\alpha)\Big]} \right] \right\}
  \right|_{\lambda = \lambda' =0}
\end{align}
where we have defined $\mu^2 = \mu_{\overline{{\text{MS}}}}^2 \,
e^{5/3}$ to make the expressions more compact.

Indeed we can not always expand the denominators of \eq{K1run} into a
geometric series employed in \eq{K1run1}, but one has to remember that
the summation of bubble chain diagrams shown on the right side of
\fig{kersub} gives one the geometric series. Hence the geometric
series come first: later they are absorbed into the denominators shown
in \eq{K1run}, which is an approximation not valid for all $q$ and
$q'$. Therefore, by keeping the geometric series in \eq{K1run1} we are
not making any approximations. In general, in what follows we are not
going to keep track of the issues of convergence of perturbation
series. The contribution of renormalons to non-linear small-$x$
evolution was thoroughly investigated in \cite{Gardi:2006rp} and was
found to be significant at low $Q^2$. We refer the interested reader
to \cite{Gardi:2006rp} for more details on this issue. 

Using the following formulas
\begin{align}\label{math1}
  \int \frac{d^2 k}{(2 \pi)^2} \, \frac{e^{- i {\un k} \cdot {\un
        z}}}{k^2 + q^2} \, = \, \frac{1}{2 \pi} \, K_0 (q \, z)
\end{align}
and
\begin{align}\label{math2}
  \int \frac{d^2 k}{(2 \pi)^2} \, e^{- i {\un k} \cdot {\un z}} \,
  \frac{{\un k}}{k^2 + q^2} \, = \, \frac{-i}{2 \pi} \, \frac{{\un
      z}}{z} \ q \, K_1 (q \, z)
\end{align}
we can now perform the $k$- and $k'$-integrals in \eq{K1run1}.
Integrating over the angles of $\un q$ and ${\un q}'$ as well yields
\begin{align}\label{K1run2}
  & {\cal K}_{\oone} ({\ud x}_0, {\ud x}_1 ; {\ud z}_1, {\ud z}_2) \,
  = \, \frac{4 \, N_f}{(2\pi)^4} \, \sum\limits_{n,m =0}^\infty (- \am
  \beta_2)^{n+m} \, \frac{d^n}{d \lambda^n} \, \frac{d^m}{d
    \lambda^{\prime m}} \, \Bigg\{ \, \int\limits_0^1 d \alpha \,
  \int\limits_0^\infty d q \, q \, d q' \, q' \left( \frac{q^2}{\mu^2}
  \right)^\lambda \, \left( \frac{q'^2}{\mu^2} \right)^{\lambda'}
  \notag \\ & \, \Bigg[ \frac{\alpha \, \abar}{{z}_{12}^2 \, |{\ud
      z}-{\ud x}_0| \, |{\ud z}-{\ud x}_1|} \, \Bigg( - 4 \, \alpha \,
  \abar \, {\ud z}_{12} \cdot ({\ud z}-{\ud x}_0) \, {\ud z}_{12}
  \cdot ({\ud z}-{\ud x}_1) + {z}_{12}^2 \, ({\ud z}-{\ud x}_0) \cdot
  ({\ud z}-{\ud x}_1) \Bigg) \notag \\ & \times \, J_1 (q \, |{\ud
    z}-{\ud x}_0|) \, K_1 (z_{12} \, q \, \sqrt{\alpha \, \abar}) \,
  J_1 (q' \, |{\ud z}-{\ud x}_1|) \, K_1 (z_{12} \, q' \, \sqrt{\alpha
    \, \abar})  + 2 \, \alpha \,
  \abar \, (\alpha - \abar) \, \sqrt{\alpha \, \abar} \notag \\
  & \times \, \Bigg( \frac{{\ud z}_{12} \cdot ({\ud z}-{\ud
      x}_0)}{{z}_{12} \, |{\ud z}-{\ud x}_0|} \, J_1 (q \, |{\ud
    z}-{\ud x}_0|) \, K_1 (z_{12} \, q \, \sqrt{\alpha \, \abar}) \,
  J_0 (q' \, |{\ud z}-{\ud x}_1|) \, K_0 (z_{12} \, q' \, \sqrt{\alpha
    \, \abar }) \notag \\ & + \, \frac{{\ud z}_{12} \cdot ({\ud
      z}-{\ud x}_1)}{{z}_{12} \, |{\ud z}-{\ud x}_1|} \, J_0 (q \,
  |{\ud z}-{\ud x}_0|) \, K_0 (z_{12} \, q \, \sqrt{\alpha \, \abar})
  \, J_1 (q' \, |{\ud z}-{\ud x}_1|) \, K_1 (z_{12} \, q' \,
  \sqrt{\alpha \, \abar}) \Bigg) \notag \\ & + 4 \, \alpha^2 \,
  \abar^2 \, J_0 (q \, |{\ud z}-{\ud x}_0|) \, K_0 (z_{12} \, q \,
  \sqrt{\alpha \, \abar}) \, J_0 (q' \, |{\ud z}-{\ud x}_1|) \, K_0
  (z_{12} \, q' \, \sqrt{\alpha \, \abar}) \Bigg] \Bigg\}
  \Bigg|_{\lambda = \lambda' =0}.
\end{align}
We have defined
\begin{align}
  \abar = 1 - \alpha
\end{align}
for brevity. Now the integrals over $q$ and $q'$ can be carried out to
give
\begin{align}\label{K1run3}
  & {\cal K}_{\oone} ({\ud x}_0, {\ud x}_1 ; {\ud z}_1, {\ud z}_2) \,
  = \, \frac{4 \, N_f}{(2\pi)^4} \, \sum\limits_{n,m =0}^\infty (- \am
  \beta_2)^{n+m} \, \frac{d^n}{d \lambda^n} \, \frac{d^m}{d
    \lambda^{\prime m}} \, \Bigg\{ \, \int\limits_0^1 d \alpha \,
  \left( \frac{4}{z_{12}^2 \, \mu^2 \, \alpha \, \abar}
  \right)^{\lambda + \lambda'} \, \Gamma^2 (1+\lambda) \notag \\ & \,
  \Gamma^2 (1+\lambda') \, \Bigg\{ \Bigg( - 4 \, \alpha \, \abar \,
  {\ud z}_{12} \cdot ({\ud z}-{\ud x}_0) \, {\ud z}_{12} \cdot ({\ud
    z}-{\ud x}_1) + {z}_{12}^2 \, ({\ud z}-{\ud x}_0) \cdot ({\ud
    z}-{\ud x}_1) \Bigg) \, \frac{(1+\lambda) \,
    (1+\lambda')}{{z}_{12}^8 \, (\alpha \, \abar)^2} \notag \\ &
  \times \, F \left( 1+\lambda, 2+\lambda; 2; - \frac{|{\ud z}-{\ud
        x}_0|^2}{\alpha \, \abar \, z_{12}^2} \right) \, F \left(
    1+\lambda', 2+\lambda'; 2; - \frac{|{\ud z}-{\ud x}_1|^2}{\alpha
      \, \abar \, z_{12}^2} \right) + 2 \,
  \frac{\alpha - \abar}{\alpha \, \abar}  \notag \\
  & \times \, \Bigg[ \frac{{\ud z}_{12} \cdot ({\ud z}-{\ud
      x}_0)}{{z}_{12}^6} \, (1+\lambda) \, F \left( 1+\lambda,
    2+\lambda; 2; - \frac{|{\ud z}-{\ud x}_0|^2}{\alpha \, \abar \,
      z_{12}^2} \right) \, F \left( 1+\lambda', 1+\lambda'; 1; -
    \frac{|{\ud z}-{\ud x}_1|^2}{\alpha \, \abar \, z_{12}^2} \right)
  \notag \\ & + \, \frac{{\ud z}_{12} \cdot ({\ud z}-{\ud
      x}_1)}{{z}_{12}^6} \, F \left( 1+\lambda, 1+\lambda; 1; -
    \frac{|{\ud z}-{\ud x}_0|^2}{\alpha \, \abar \, z_{12}^2} \right)
  \, (1+\lambda') \, F \left( 1+\lambda', 2+\lambda'; 2; - \frac{|{\ud
        z}-{\ud x}_1|^2}{\alpha \, \abar \, z_{12}^2} \right) \Bigg]
  \notag \\ & + \frac{4}{z_{12}^4} \, F \left( 1+\lambda, 1+\lambda;
    1; - \frac{|{\ud z}-{\ud x}_0|^2}{\alpha \, \abar \, z_{12}^2}
  \right) \, F \left( 1+\lambda', 1+\lambda'; 1; - \frac{|{\ud z}-{\ud
        x}_1|^2}{\alpha \, \abar \, z_{12}^2} \right) \Bigg\} \Bigg\}
  \Bigg|_{\lambda = \lambda' =0}.
\end{align}
Unfortunately further simplification of the expression in \eq{K1run3}
is impossible without approximations. The series resulting from
summation over $n$ and $m$ are likely to be divergent due to
renormalons. As we mentioned before, here we neglect the renormalon
problem referring the reader to \cite{Gardi:2006rp}. Similar to how it
was done in \cite{Kovchegov:2006vj} we are not going to attempt to
resum the series exactly: instead we will calculate the
next-to-leading order terms and assume that with a good accuracy they
give us the scale(s) of the running coupling constant. This procedure
is similar to the well-known prescription due to Brodsky, Lepage and
Mackenzie \cite{Brodsky:1983gc}.

Using the Taylor-expansions of hypergeometric functions
\begin{align}\label{B7}
  F \left( 1+\lambda, 2+\lambda; 2; z \right) \, = \,
  \frac{1}{1-z} - \lambda \, \frac{1}{1-z} \, \left[ 1 + \ln (1-z) +
    \frac{1}{z} \, \ln (1-z) \right] + o(\lambda^2).
\end{align}
and
\begin{align}
  F \left( 1+\lambda, 1+\lambda; 1; z \right) \, = \, \frac{1}{1-z} -
  \lambda \, \frac{2}{1-z} \, \ln \left( 1-z \right) + o(\lambda^2)
\end{align}
after some algebra we obtain
\begin{align}\label{K1run4}
  & {\cal K}_{\oone} ({\ud x}_0, {\ud x}_1 ; {\ud z}_1, {\ud z}_2) \,
  = \, \frac{N_f}{4 \pi^4} \, \int\limits_0^1 d \alpha \,
  \frac{1}{[\alpha \, ({\un z}_1 - {\un x}_0)^2 + \abar \, ({\un z}_2
    - {\un x}_0)^2] \, [\alpha \, ({\un z}_1 - {\un x}_1)^2 + \abar \,
    ({\un z}_2 - {\un x}_0)^2] \, z_{12}^4} \notag \\ & \Bigg\{ \Bigg[
  - 4 \, \alpha \, \abar \, {\ud z}_{12} \cdot ({\ud z}-{\ud x}_0) \,
  {\ud z}_{12} \cdot ({\ud z}-{\ud x}_1) + {z}_{12}^2 \, ({\ud z}-{\ud
    x}_0) \cdot ({\ud z}-{\ud x}_1) \Bigg] \notag \\ \displaybreak[0]
  & \times \, \left[ 1 - \am \, \beta_2 \, \ln \left( \frac{1}{R_T^2
        ({\un x}_0) \, \mu_{\overline{{\text{MS}}}}^2} \right) + o
    (\am^2) \right] \, \left[ 1 - \am \, \beta_2 \, \ln \left(
      \frac{1}{R_T^2 ({\un x}_1) \, \mu_{\overline{{\text{MS}}}}^2}
    \right) + o (\am^2) \right] \notag \\ & + 2 \, \alpha \, \abar \,
  (\alpha - \abar) \, z_{12}^2 \, \left\{ {\ud z}_{12} \cdot ({\ud
      z}-{\ud x}_0) \, \left[ 1 - \am \, \beta_2 \, \ln \left(
        \frac{1}{R_T^2 ({\un x}_0) \, \mu_{\overline{{\text{MS}}}}^2}
      \right) + o (\am^2) \right] \right. \notag \\ & \times \, \left[
    1 - \am \, \beta_2 \, \ln \left( \frac{1}{R_L^2 ({\un x}_1) \,
        \mu_{\overline{{\text{MS}}}}^2} \right) + o (\am^2) \right] +
  {\ud z}_{12} \cdot ({\ud z}-{\ud x}_1) \notag \\ & \left. \times \,
    \left[ 1 - \am \, \beta_2 \, \ln \left( \frac{1}{R_L^2 ({\un x}_0)
          \, \mu_{\overline{{\text{MS}}}}^2} \right) + o (\am^2)
    \right] \, \left[ 1 - \am \, \beta_2 \, \ln \left( \frac{1}{R_T^2
          ({\un x}_1) \, \mu_{\overline{{\text{MS}}}}^2} \right) + o
      (\am^2) \right] \right\} \notag \\ & \left. + 4 \, \alpha^2 \,
    \abar^2 \, z_{12}^4 \, \left[ 1 - \am \, \beta_2 \, \ln \left(
        \frac{1}{R_L^2 ({\un x}_0) \, \mu_{\overline{{\text{MS}}}}^2}
      \right) + o (\am^2) \right] \, \left[ 1 - \am \, \beta_2 \, \ln
      \left( \frac{1}{R_L^2 ({\un x}_1) \,
          \mu_{\overline{{\text{MS}}}}^2} \right) + o (\am^2) \right]
  \right\}.
\end{align}
In arriving at \eq{K1run4} we employed functions $R_T ({\un x})$ and
$R_L ({\un x})$, which have dimensions of transverse coordinates and
are defined by
\begin{align}\label{RT}
  \ln \left( \frac{1}{R_T^2 ({\un x}) \,
      \mu_{\overline{{\text{MS}}}}^2} \right) \, = \, & \ln \left(
    \frac{4 \, e^{- 2 \gamma - 5/3}}{[\alpha \, ({\un z}_1 - {\un
        x})^2 + \abar \, ({\un z}_2 - {\un x})^2] \,
      \mu_{\overline{{\text{MS}}}}^2} \right) \notag \\ & +
  \frac{\alpha \, \abar \, z_{12}^2}{({\un z} - {\un x})^2} \, \ln
  \left( \frac{\alpha \, ({\un z}_1 - {\un x})^2 + \abar \, ({\un z}_2
      - {\un x})^2}{\alpha \, \abar \, z_{12}^2} \right)
\end{align}
and
\begin{align}\label{RL}
  \ln \left( \frac{1}{R_L^2 ({\un x}) \,
      \mu_{\overline{{\text{MS}}}}^2} \right) \, = \, \ln \left(
    \frac{4 \, e^{- 2 \gamma - 5/3}}{[\alpha \, ({\un z}_1 - {\un
        x})^2 + \abar \, ({\un z}_2 - {\un x})^2] \,
      \mu_{\overline{{\text{MS}}}}^2} \right) - \ln \left(
    \frac{\alpha \, ({\un z}_1 - {\un x})^2 + \abar \, ({\un z}_2 -
      {\un x})^2}{\alpha \, \abar \, z_{12}^2} \right).
\end{align}
The subscripts T and L stand for transverse and longitudinal gluon
polarizations which give rise to the two different functions under the
logarithm.

Recombining the series in \eq{K1run4} into physical running couplings
finally yields
\begin{align}\label{K1run5}
  & \am^2 \, {\cal K}_{\oone} ({\ud x}_0, {\ud x}_1 ; {\ud z}_1, {\ud
    z}_2) \, = \, \frac{N_f}{4 \pi^4} \, \int\limits_0^1 d \alpha \,
  \frac{1}{[\alpha \, ({\un z}_1 - {\un x}_0)^2 + \abar \, ({\un z}_2
    - {\un x}_0)^2] \, [\alpha \, ({\un z}_1 - {\un x}_1)^2 + \abar \,
    ({\un z}_2 - {\un x}_0)^2] \, z_{12}^4} \notag \\ & \Bigg\{ \Bigg[
  - 4 \, \alpha \, \abar \, {\ud z}_{12} \cdot ({\ud z}-{\ud x}_0) \,
  {\ud z}_{12} \cdot ({\ud z}-{\ud x}_1) + {z}_{12}^2 \, ({\ud z}-{\ud
    x}_0) \cdot ({\ud z}-{\ud x}_1) \Bigg] \, \as \left(
    \frac{1}{R_T^2 ({\un x}_0)} \right) \, \as \left( \frac{1}{R_T^2
      ({\un x}_1)} \right) \notag \\ & + 2 \, \alpha \, \abar \,
  (\alpha - \abar) \, z_{12}^2 \, \left[ {\ud z}_{12} \cdot ({\ud
      z}-{\ud x}_0) \, \, \as \left( \frac{1}{R_T^2 ({\un x}_0)}
    \right) \, \as \left( \frac{1}{R_L^2 ({\un x}_1)} \right) \right.
  \notag \\ & \left.  \left. + {\ud z}_{12} \cdot ({\ud z}-{\ud x}_1)
      \, \as \left( \frac{1}{R_L^2 ({\un x}_0)} \right) \, \as \left(
        \frac{1}{R_T^2 ({\un x}_1)} \right) \right] + 4 \, \alpha^2 \,
    \abar^2 \, z_{12}^4 \, \as \left( \frac{1}{R_L^2 ({\un x}_0)}
    \right) \, \as \left( \frac{1}{R_L^2 ({\un x}_1)} \right)
  \right\}
\end{align}
with the physical running coupling in the $\overline{{\text{MS}}}$
scheme given by
\begin{align}
  \as (1/R^2) \, = \, \frac{\am}{ 1 + \am \beta_2 \, \ln \left(
      \frac{1}{R^2 \, \mu_{\overline{{\text{MS}}}}^2} \right)}.
\label{alpha}
\end{align}

\eq{K1run5} is the contribution to the JIMWLK evolution kernel of the
resummed diagram on the right hand side of \fig{kersub}.

%%%%%%%%%%%%%%%%%%%%%%%%%%%%%%%%%%%%%%%%%%%%%%%%%%%%%%%%%%%%%%%%%%%%%%%%

\subsection{Brief summary of analytical results}
\label{analyt_sum}

Let us briefly summarize our analytical results. The non-linear
small-$x$ evolution equation with the running coupling corrections
included reads
\begin{align}
  \frac{\partial S(\ud{x}_0,\ud{x}_1;Y)}{\partial Y} \, = \,
  \mathcal{R}\left[S\right]-\mathcal{S}\left[S\right]\,.
\label{frs}
\end{align}

The first term on the right hand side of \eq{frs} is referred to as
the running coupling contribution. It was calculated independently in
\cite{Balitsky:2006wa} and in \cite{Kovchegov:2006vj}: the results of
those calculations are given above in Eqs. (\ref{kbal}) and
(\ref{kkw}) correspondingly, which have to be combined with \eq{run}
to obtain 
\begin{align}\label{bal_run}
  \mathcal{R}^{\text{Bal}} \left[S\right] \, = \, \int d^2 z \,
  \tilde{K}^{\text{Bal}} (\ud{x}_0,\ud{x}_1, {\un z})
  \left[S(\ud{x}_0,\ud{z};Y)\,S(\ud{z},\ud{x}_1;Y)-S(\ud{x}_0,\ud{x}_1;Y)\right]
\end{align}
and
\begin{align}\label{kw_run}
  \mathcal{R}^{\text{KW}} \left[S\right] \, = \, \int d^2 z \,
  \tilde{K}^{\text{KW}} (\ud{x}_0,\ud{x}_1, {\un z})
  \left[S(\ud{x}_0,\ud{z};Y)\,S(\ud{z},\ud{x}_1;Y)-S(\ud{x}_0,\ud{x}_1;Y)\right].
\end{align}
One notices immediately that $\mathcal{R}^{\text{Bal}} \left[S\right]$
calculated in \cite{Balitsky:2006wa} is different from
$\mathcal{R}^{\text{KW}} \left[S\right]$ calculated in
\cite{Kovchegov:2006vj} due to the difference in the kernels
$\tilde{K}^{\text{Bal}}$ and $\tilde{K}^{\text{KW}}$ in Eqs.
(\ref{kbal}) and (\ref{kkw}). However that does not imply disagreement
between the calculations of \cite{Balitsky:2006wa} and
\cite{Kovchegov:2006vj}: after all, it is the full kernel on the right
of \eq{frs}, $\mathcal{R}\left[S\right]-\mathcal{S}\left[S\right]$,
that needs to be compared. To do that one has to calculate the second
term on the right hand side of \eq{frs}.

The second term on the right hand side of \eq{frs} is referred to as
the subtraction contribution. It is given by 
\begin{align}\label{sub_full}
  {\mathcal S} [S] \, = \, \am^2 \, \int d^2 z_1 \, d^2 z_2 \,
  K_{\oone} ({\un x}_0, {\un x}_1 ; {\un z}_1, {\un z}_2) \, [ S ({\un
    x}_{0}, {\un w}, Y) \, S ({\un w}, {\un x}_1, Y) - S ({\un x}_{0},
  {\un z}_1, Y) \, S ({\un z}_{2}, {\un x}_1, Y)]
\end{align}
with the resummed BK kernel
\begin{align}\label{fullBK}
  K_{\oone} ({\un x}_0, {\un x}_1 ; {\un z}_1, {\un z}_2) \, = \, C_F
  \, \sum_{m,n = 0}^1 \, (-1)^{m+n} \, {\cal K}_{\oone} & ({\un x}_m,
  {\un x}_n ; {\un z}_1, {\un z}_2).
\end{align}
The resummed JIMWLK kernel ${\cal K}_{\oone} ({\un x}_m, {\un x}_n ;
{\un z}_1, {\un z}_2)$ is given by \eq{K1run5}, along with Eqs.
(\ref{RT}) and (\ref{RL}) defining the scales of the running
couplings. In the numerical solution below we will replace $N_f
\rightarrow - 6 \pi \beta_2$ in its prefactor, obtaining
\begin{align}\label{K1run_final}
  & \am^2 \, {\cal K}_{\oone} ({\ud x}_0, {\ud x}_1 ; {\ud z}_1, {\ud
    z}_2) = - \frac{3 \, \beta_2}{2 \pi^3} \hspace*{-.5mm}
  \int\limits_0^1 \hspace*{-.5mm} d \alpha \, \frac{1}{[\alpha \,
    ({\un z}_1 - {\un x}_0)^2 + \abar \, ({\un z}_2 - {\un x}_0)^2] \,
    [\alpha \, ({\un z}_1 - {\un x}_1)^2 + \abar \, ({\un z}_2 - {\un
      x}_0)^2] \, z_{12}^4} \notag \\ & \Bigg\{ \Bigg[ - 4 \, \alpha
  \, \abar \, {\ud z}_{12} \cdot ({\ud z}-{\ud x}_0) \, {\ud z}_{12}
  \cdot ({\ud z}-{\ud x}_1) + {z}_{12}^2 \, ({\ud z}-{\ud x}_0) \cdot
  ({\ud z}-{\ud x}_1) \Bigg] \, \as \left( \frac{1}{R_T^2 ({\un x}_0)}
  \right) \, \as \left( \frac{1}{R_T^2 ({\un x}_1)} \right) \notag \\ 
  & + 2 \, \alpha \, \abar \, (\alpha - \abar) \, z_{12}^2 \, \left[
    {\ud z}_{12} \cdot ({\ud z}-{\ud x}_0) \, \, \as \left(
      \frac{1}{R_T^2 ({\un x}_0)} \right) \, \as \left( \frac{1}{R_L^2
        ({\un x}_1)} \right) \right.  \notag \\ & \left.  \left. +
      {\ud z}_{12} \cdot ({\ud z}-{\ud x}_1) \, \as \left(
        \frac{1}{R_L^2 ({\un x}_0)} \right) \, \as \left(
        \frac{1}{R_T^2 ({\un x}_1)} \right) \right] + 4 \, \alpha^2 \,
    \abar^2 \, z_{12}^4 \, \as \left( \frac{1}{R_L^2 ({\un x}_0)}
    \right) \, \as \left( \frac{1}{R_L^2 ({\un x}_1)} \right) \right\}.
\end{align}
This substitution is the same as for all other factors of $N_f$. The
same substitution was performed in \cite{Kovchegov:2006vj} to
calculate the running coupling term. In fact, as was shown in
\cite{Kovchegov:2006wf}, the linear part of the subtraction term
(calculated using the prescription of \cite{Kovchegov:2006vj})
contributes to the running coupling corrections to the BFKL equation.
Therefore, in that case, the factor of $N_f$ in front of \eq{K1run5}
is definitely a part of the beta-function.  Hence the replacement $N_f
\rightarrow - 6 \pi \beta_2$ is justified even in the subtraction
term. Once again, in the numerical solution below we will use
\eq{K1run_final} along with \eq{fullBK} in \eq{sub_full} to calculate
the subtraction term ${\mathcal S} [S]$.

Substituting ${\un w} = {\un z}_1$ (or, equivalently, ${\un w} = {\un
  z}_2$) in \eq{sub_full} would yield the subtraction term
\begin{align}\label{sub_full_bal}
  {\mathcal S}^{\text{Bal}} [S] \, = \, & \am^2 \, \int d^2 z_1 \, d^2
  z_2 \, K_{\oone} ({\un x}_0, {\un x}_1 ; {\un z}_1, {\un z}_2)
  \notag \\ & \times \, [ S ({\un x}_{0}, {\un z}_1, Y) \, S ({\un
    z}_1, {\un x}_1, Y) - S ({\un x}_{0}, {\un z}_1, Y) \, S ({\un
    z}_{2}, {\un x}_1, Y)]
\end{align}
which has to be subtracted from $\mathcal{R}^{\text{Bal}}
\left[S\right]$ calculated in \cite{Balitsky:2006wa} and given by
\eq{bal_run} to obtain the complete evolution equation resumming all
orders of $\as \, N_f$ in the kernel. 

Substituting ${\un w} = {\un z} = \alpha \, {\un z}_1 + (1-\alpha) \,
{\un z}_2$ in \eq{sub_full} yields
\begin{align}\label{sub_full_kw}
  {\mathcal S}^{\text{KW}} [S] \, = \, & \am^2 \, \int d^2 z_1 \, d^2
  z_2 \, K_{\oone} ({\un x}_0, {\un x}_1 ; {\un z}_1, {\un z}_2)
  \notag \\ & \times \, [ S ({\un x}_{0}, {\un z}, Y) \, S ({\un z},
  {\un x}_1, Y) - S ({\un x}_{0}, {\un z}_1, Y) \, S ({\un z}_{2},
  {\un x}_1, Y)]
\end{align}
which has to be subtracted from $\mathcal{R}^{\text{KW}}
\left[S\right]$ calculated in \cite{Kovchegov:2006vj} and given in
\eq{kw_run} again to obtain the complete evolution equation resumming
all orders of $\as \, N_f$ in the kernel. We checked explicitly by
performing analytic calculations that the two evolution equations
obtained this way agree at the NLO and NNLO. Below we will check the
agreement of the two calculations to all orders by performing a
numerical analysis of the solutions of these equations.

The above discussion demonstrates that the separation of the evolution
kernel into the running coupling and subtraction pieces, as done in
\eq{frs}, is somewhat artificial, and has no small parameter
justifying one or another separation prescription. Therefore, the
small-$x$ evolution equation including all running coupling (or, more
precisely, $\as \, N_f$) corrections should combine both terms in
\eq{frs}. Below we will solve such evolution equation numerically to
obtain the full small-$x$ evolution with the running coupling.

%%%%%%%%%%%%%%%%%%%%%%%%%%%%%%%%%%%%%%%%%%%%%%%%%%%%%%%%%%%%%%%%%%%%%%%%

\section{Numerical setup and initial conditions} \label{setup}

In our numerical study we consider the translational invariant
approximation in which the scattering matrix is independent of the
impact parameter of the collision, i.e., $S=S(r,Y)$.  To solve the
integro-differential equations, corresponding to the BK equation with
running coupling we employ a second-order Runge-Kutta method with a
step size in rapidity $\Delta Y=0.1$. We discretize the variable
$|\ud{r}|$ into 800 points equally separated in logarithmic space
between $r_{min}= 10^{-8}$ and $r_{max}=50$.  Throughout this paper,
the units of $r$ will be GeV$^{-1}$, and those of $Q_s$ will be GeV.
All the integrals have been performed using improved adaptative
Gaussian quadrature methods.  The accuracy of this numerical method
has been checked in \cite{Albacete:2004gw} to be better than a $4\%$
in all the $r$ range.

We consider three different initial conditions for the dipole
scattering amplitude, $N(r,Y)=1-S(r,Y)$.  The first one is taken from
the McLerran-Venugopalan (MV) model
\cite{McLerran:1993ni,McLerran:1993ka}:
\begin{align}
  N^{MV}(r,Y=0)=1-\exp{\left[-\frac{r^2Q_s^{'2}}{4}
\ln{\left(\frac{1}{r^2\Lambda^2}+e\right)}\right]}\, .
\label{mv}
\end{align}
where a constant term has been added to the argument of the logarithm
in the exponent in order to regularize it for large values of $r$. The
other two initial conditions are given by
\begin{align}
  N^{AN}(r)=1-\exp{\left[-\frac{(r\,Q_s')^{2\gamma}}{4}\right]}\, ,
\label{an}
\end{align}
with $\gamma=0.6$ and $\gamma=0.8$. These two last initial conditions
will be referred hereinafter as AN06 and AN08 respectively.  The
interest in this ansatz, reminiscent of the Golec-Biernat--Wusthoff 
model \cite{{Golec-Biernat:1998js}}, is that the small-$r$
behavior $N^{AN} \propto r^{2\gamma}$ corresponds to an anomalous
dimension $1-\gamma$ of the unintegrated gluon distribution at large
transverse momentum.  (AN labels initial conditions with anomalous
dimension.)  Our choices $\gamma=0.6$ and $\gamma=0.8$ can be
motivated a posteriori by the observation that the anomalous dimension
of the evolved BK solution for running coupling lies in between those
two values and the one for the MV initial condition, $\gamma\approx 1$
(see Section \ref{scaling}). Thus, the choice of distinct initial
conditions allows us to better track the onset of the expected
asymptotic universal behavior that is eventually reached at high
energies and to study the influence of the pre-asymptotic,
non-universal corrections to the solutions of the evolution equations.
To completely determine our initial conditions, we set $Q_s'=1$~GeV at
Y=0 in Eqs. (\ref{mv}) and (\ref{an}) and put $\Lambda = 0.2$~GeV.
Although $Q_s'$ is normally identified with the saturation scale, our
definition of the saturation scale through the rest of the paper will
be purely pragmatical and given by the condition
\begin{align}
  N(r=1/Q_s(Y), Y)=\kappa,
\label{qs}
\end{align}
with $\kappa=0.5$. We have checked that this choice of $\kappa$,
albeit arbitrary, does not affect any of the major conclusions to be
drawn in the rest of the paper.

Finally, in order to avoid the Landau pole and to regularize the
running coupling at large transverse sizes we stick to the following
procedure: for small transverse distances $r< r_{fr}$, with $r_{fr}$
defined by $\alpha_s(1/r_{fr}^2)=0.5$, the running coupling is given
by the one loop expression
\begin{align}
  \as (1/r^2) \, = \, \frac{1}{\beta_2 \, \ln \left( \frac{1}{r^2 \,
        \Lambda^2 }\right)}
\end{align}
with $N_f=3$ and $\Lambda = 0.2$~GeV, whereas for larger sizes,
$r>r_{fr}$, we ``freeze'' the coupling at a fixed value
$\alpha_s=0.5$. A detailed study of the role of Landau pole in
non-linear small-$x$ evolution is given in \cite{Gardi:2006rp}.

%%%%%%%%%%%%%%%%%%%%%%%%%%%%%%%%%%%%%%%%%%%%%%%%%%%%%%%%%%%%%%%%%%%%%%%%%%

\section{Results} \label{results}

In this Section, we discuss our numerical results and how they compare
to previous numerical work and analytical estimates.

%%%%%%%%%%%%%%%%%%%%%%%%%%%%%%%%%%%%%%%%%%%%%%%%%%%%%%%%%%%%%%%%%%

\subsection{Running coupling}\label{running}

\begin{figure}[hbt]
\begin{center}
\includegraphics[height=13.25cm]{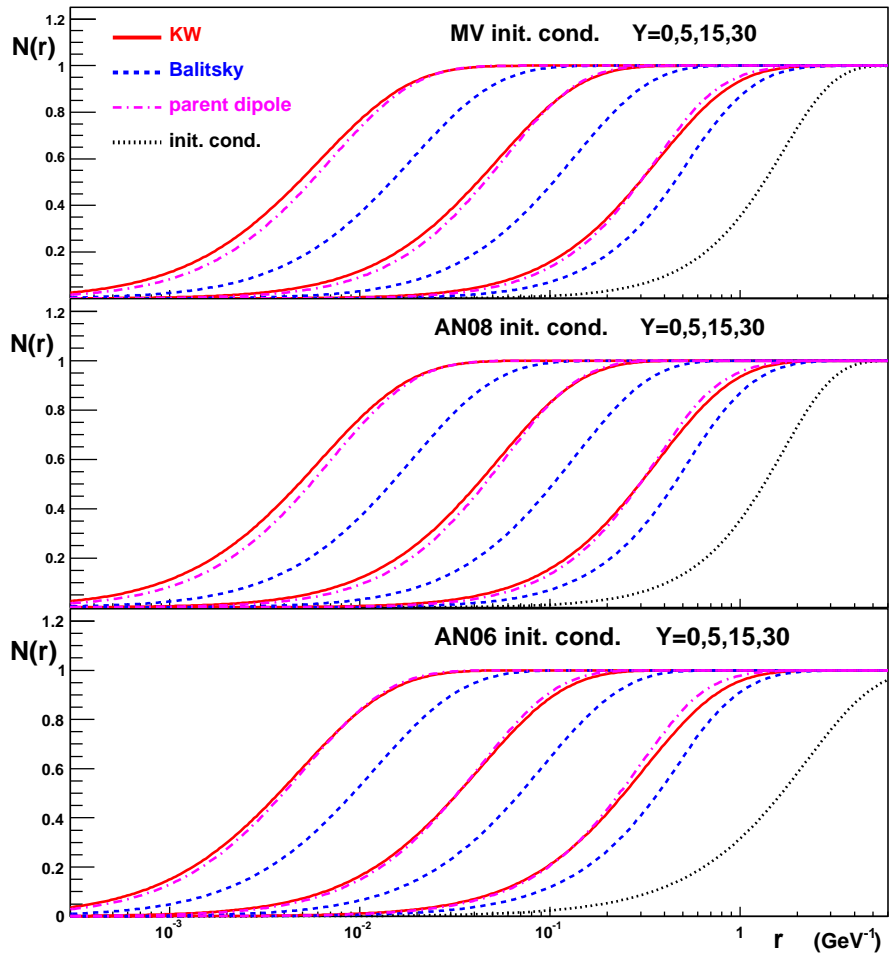}%,width=14.5cm]{c1.eps}
\caption{Solutions of the BK equation at rapidities Y=0, 5, 15 and 30
  (curves are labeled from right to left) for the three running
  coupling schemes considered in this work: KW (solid line), Balitsky
  (dashed line) and parent dipole (dashed-dotted lines). The initial
  conditions are MV (top), AN08 (middle) and AN06 (bottom).}
\label{sols}
\end{center}
\end{figure}

\fig{sols} shows the solutions of the evolution equation when only the
running coupling contribution is taken into account, i.e., neglecting
the subtraction term in \eq{frs}, for different initial conditions and
for the three schemes considered in this work: Balitsky's, given by
Eqs. (\ref{kbal}) and (\ref{bal_run}), KW, given by Eqs. (\ref{kkw})
and (\ref{kw_run}), and the the ad hoc parent dipole implementation of
the running coupling, shown in \eq{kpd}.

%%%%%%%%%%%%%%%%%%%%%%%%%%%%%
%%%%%%%%%%%%%%%%%%%%%%%%%

As previously observed in \cite{Albacete:2004gw,Braun:2003un}, the
most relevant effect of including running coupling corrections in the
evolution equation is a considerable reduction in the speed of the
evolution with respect to the fixed coupling case. This is a common
feature of the different running coupling schemes studied here and of
other phenomenological ones considered in the literature (a detailed
comparison between the solutions for fixed coupling evolution and for
parent dipole running coupling can be found e.g. in
\cite{Albacete:2004gw}).  This is not a surprising result, since a
generic effect of the running of the coupling is to suppress the
emission of small transverse size dipoles, which is the leading
mechanism driving the evolution.

However, despite this common feature of the running coupling
solutions, significant differences are found between the solutions
obtained under different schemes as we infer from \fig{sols}. In
particular, the evolution is much faster with the KW prescription than
with that of Balitsky.  Equivalently, the KW prescription yields a
stronger growth of the saturation scale with rapidity/energy than
Balitsky's. Moreover, the solutions obtained when the parent dipole
prescription is used lay much closer to those obtained within the KW
scheme than to the ones obtained when Balitsky's scheme is applied,
contrary to what was suggested in \cite{Balitsky:2006wa}. As argued
before, the differences observed in the solutions obtained using the
two subtraction schemes are entirely due to neglecting the subtraction
contribution and reflect the arbitrariness of the separation
procedure.

%%%%%%%%%%%%%%%%%%%%%%%%%%%%%%%%%%%%%%%%%%%%%%%%%%%%%%%%%%%%%%%%%%%%%%%%

\subsection{Geometric scaling}\label{scaling}

It has been found in previous analytical
\cite{Iancu:2002tr,Mueller:2002zm,Munier:2003vc} and numerical studies
on the solutions of the BK equation at leading order
\cite{Albacete:2004gw,Albacete:2003iq,
  Lublinsky:2001bc,Armesto:2001fa} and for different heuristic
implementations of next-to-leading order corrections
\cite{Albacete:2004gw, Braun:2003un}, including the parent dipole
prescription for the running coupling also considered in this work,
that the solutions of the evolution equation at high enough rapidities
are no longer a function of two separate variables $r$ and $Y$, but
rather they depend on a single scaling variable, $\tau=r\,Q_s(Y)$.
This feature of the evolution, commonly referred to as {\it geometric
  scaling}, is an exact property of the solutions for fixed coupling
evolution due to the conformal invariance of the leading-log kernel,
and has become one of the key connections between the saturation based
formalisms and the phenomenology of heavy ion collisions and deep
inelastic scattering experiments
\cite{Stasto:2000er,Armesto:2004ud,Albacete:2005ef,Iancu:2003ge,Dumitru:2005kb,Goncalves:2006yt,Kharzeev:2004yx}.

\begin{figure}[ht]
\begin{center}
\includegraphics[width=17cm]{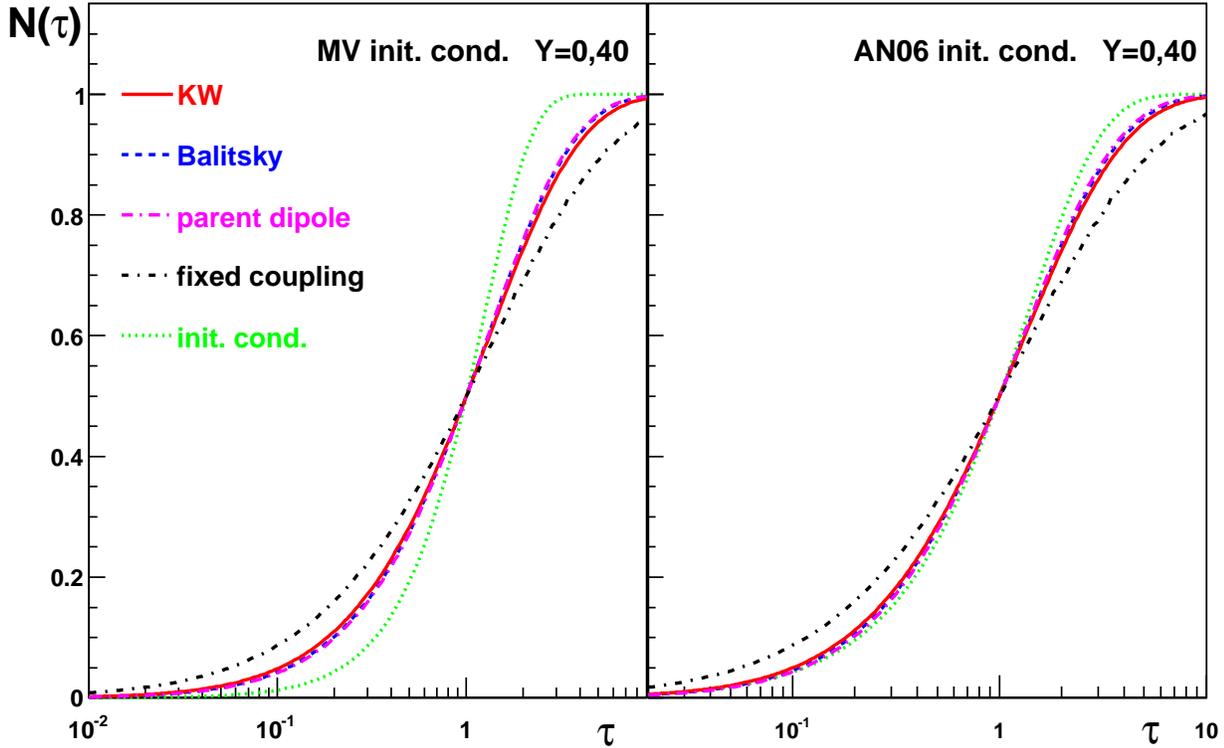}%,width=14.5cm]{c1.eps}
\caption{Solutions of the BK equation at rapidities Y=0 and 40
  for KW (solid line), Balitsky (dashed line) and parent dipole
  (dashed-dotted lines) schemes plotted versus the scaling variable
  $\tau=rQ_s(Y)$. The asymptotic solution obtained with fixed coupling
  $\alpha_s=0.2$ at $Y=40$ in \cite{Albacete:2004gw} is shown (black
  dashed-dotted line) for comparison. The initial conditions are MV
  (left) and AN06 (right). }
\label{scal}
\end{center}
\end{figure}

It can be seen from \fig{scal} that the solutions of the BK equation
with the running coupling terms discussed in the previous section also
exhibit the property of scaling, in agreement with the analytical
study carried out in \cite{Beuf:2007cw}, shown by the fact that the
rescaled high rapidity solutions lay on a single curve which is
independent of both the running coupling scheme and of the initial
condition. The scaling behavior of the solution is observed in the
whole $\tau$ range studied in this work, including the saturation
region, $\tau>1$.  The tiny deviations from a pure scaling behavior
observed in \fig{scal} may be attributed to the fact that the full
asymptotic behavior is reached at even larger rapidities ($Y\gtrsim
80$, \cite{Albacete:2004gw}) than those achieved by the numerical
solution performed in this work.
%%%%%%%%%%%%%%%%%%%%%%%%%%%%%
%%%%%%%%%%%%%%%%%%%%%%%%%%%%%

Remarkably, the scaling function for both KW and Balitsky's scheme
coincides with the one obtained with the parent dipole prescription,
up to the above mentioned scaling violations. It has been observed in
\cite{Albacete:2004gw,Albacete:2003iq,Braun:2003un} that the scaling
function differs significantly in the fixed and running coupling
cases. Following that work, and to make a more quantitative study of
the scaling property, we fitted our solutions to the functional form
\cite{Mueller:2002zm} 
\begin{align}
  f(\tau)=a\, \tau^{2\,\gamma}\left(\,\ln
  \tau^2 +b\,\right)\, ,
\label{scalfun}
\end{align}
with $a$, $b$ and $\gamma$ free parameters, within a fixed window
below the saturation region, $\tau\in[10^{-5},0.1]$. Noticeably, at
large enough rapidities the whole fitting window lays within the
geometric scaling window proposed in \cite{Iancu:2002tr}:
$(\Lambda/Q_s(Y))<\tau<1$, where $\Lambda$ is some initial scale.  The
value of $\gamma$ extracted from the fits at rapidity $Y=40$ lays in
between $\gamma\sim 0.8$ and $\gamma\sim 0.9$. This conclusion holds
for the three initial conditions used here: the anomalous dimension
seems to converge to some intermediate value, in agreement with the
value found in \cite{Albacete:2004gw}, for asymptotic running coupling
solutions ($\gamma\sim 0.85$ at $Y=70$).  This result for anomalous
dimension is very far away from the value obtained in
\cite{Albacete:2004gw} for fixed coupling solutions ($\gamma\sim 0.64$
at $Y=70$) and from the predicted anomalous dimension for both running
and fixed coupling solutions from analytical studies of the equation
based on saddle point techniques
\cite{Mueller:2002zm,Triantafyllopoulos:2002nz,Iancu:2002tr,Munier:2004xu,Beuf:2007cw},
$\gamma_c=\chi(\gamma_c)/\chi'(\gamma_c)=0.6275$, where $\chi$ is the
leading-log BFKL kernel.

%%%%%%%%%%%%%%%%%%%%%%%%%%%%%
\begin{figure}[ht]
\begin{center}
\includegraphics[height=9cm]{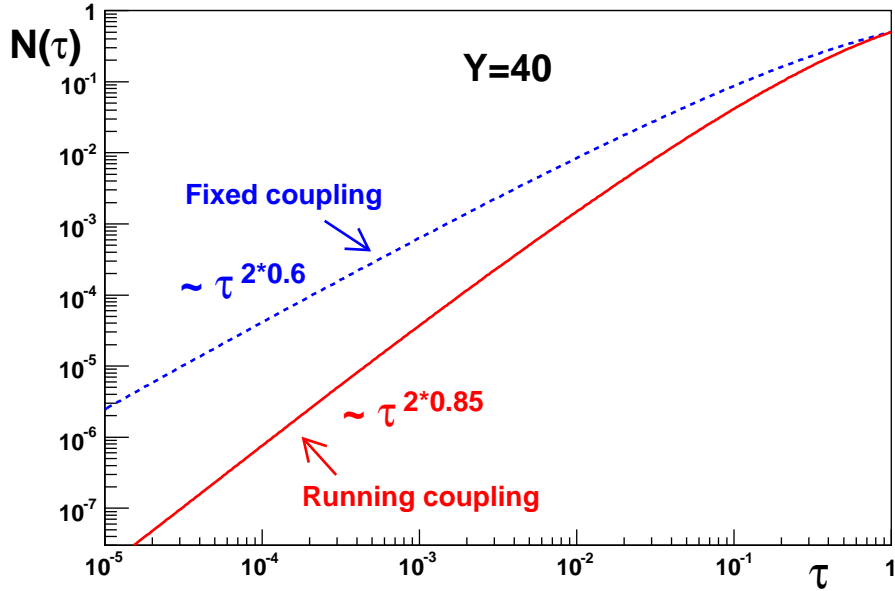}%,width=14.5cm]{c1.eps}
\caption{Asymptotic solutions (Y=40) of the evolution equation for
  running coupling (solid line) and fixed coupling with $\alpha_s=0.2$
  (dashed line). A fit to a power-law function $a\tau^{2\gamma}$ in
  the region $\tau\in[10^{-6},10^{-2}]$ yields $\gamma\approx 0.85$
  for the running coupling solution and $\gamma\approx 0.6$ for the
  fixed coupling one.}
\label{anom}
\end{center}
\end{figure}
%%%%%%%%%%%%%%%%%%%%%%%%%%%%%

It might be argued that the numerical value of the anomalous dimension
extracted from our fits is conditioned by the choice of the fitting
function and by the fitting interval. Actually, it was shown in
\cite{Albacete:2004gw} that the solutions of the evolution could be
well fitted by other functional forms, including the
double-leading-log solution of BFKL, within a similar fitting region
to the one considered in this work. On the other hand, several
phenomenological parameterizations of the solution of the evolution
have been proposed in
\cite{Iancu:2003ge,Dumitru:2005kb,Goncalves:2006yt,Kharzeev:2004yx}
and have successfully confronted HERA and RHIC experimental data.
There, the dipole scattering amplitude at arbitrary rapidity is
assumed to be given by a functional form analogous to our ansatz for
the initial condition \eq{an}, but allowing for geometric scaling
violations by replacing $\gamma\rightarrow\gamma(r,Y)$. The value of
the anomalous dimension at $r=1/Q_s$ and/or for $Y\rightarrow \infty$
is fixed to be the BFKL saddle point, $\gamma_c\sim 0.63$ (the saddle
point value considered in \cite{Kharzeev:2004yx} is slightly
different, $\gamma\sim 0.53$), while the value $\gamma=1$ is recovered
in the limit $r\rightarrow \infty$ at any finite rapidity.  The
success of these phenomenological works supports the claim that the
anomalous dimension of the solution is given by the BFKL saddle point,
in agreement with the above mentioned analytical predictions.
However, the relevant values of momenta probed at current
phenomenological applications are very distinct from the fitting
region considered here. For example, the inclusive structure function
measured in HERA is fitted in \cite{Iancu:2003ge,Goncalves:2006yt}
within the region $0.045$~GeV$^2$~$<Q^2<45$~GeV$^2$, whereas charged
hadron $p_t$ spectra in dAu collisions is well reproduced by
\cite{Dumitru:2005kb,Kharzeev:2004yx,Goncalves:2006yt} in the region
$1$~GeV~$<p_t<4.5$~GeV. Note that, for both sets of data, the measured
regions overlap with the deeply saturated domain of the solution. On
the contrary, our fitting region $10^{-5}<\tau<1$ corresponds to
values of momenta $\sim 10\,Q_s(Y)<p_t<10^5\, Q_s(Y)$ (always well
above the saturation scale), with $Q_s(Y=40)\sim 500\div 1000$ GeV for
the different running coupling schemes considered and, therefore, has
no overlap with the kinematic regions measured experimentally, since
we scrutinize a momentum region strongly shifted to the ultraviolet
compared to currently available data. Moreover, it should be noticed
that the rapidity interval covered by both sets of experimental data is
$\Delta Y<4$ in both cases, while we study the solutions of the
evolution at asymptotic rapidities, $Y\sim 40$. We have checked that
shifting our fitting region to larger values of $\tau$ (smaller
momentum) would bring the value of $\gamma$ extracted from our fits
closer to the saddle point BFKL one, since the transition from the
ultraviolet region to the deeply saturated domain of the scaling
solution is realized by a locally less steeper function (see Figs.
(\ref{scal}) and (\ref{scalsub})). Therefore, there is no
contradiction at all between the success of the phenomenological
parameterizations of the solutions and the results reported here.

With the above clarifications we reach the following conclusion: the
asymptotic scaling solutions corresponding to fixed and running
coupling evolution are intrinsically different in the whole $r$-range.
This is emphasized in \fig{anom}, where we represent the scaling
solutions in a log scale for $\tau<1$. It is clear that the tail of
the distribution falls off with decreasing $\tau$ much steeper for the
running coupling solution than for the fixed coupling one. A fit to a
pure power-law function, $f=a\,\tau^{2\gamma}$, in the region
$\tau\in[10^{-6},10^{-2}]$ yields $\gamma\sim0.85$ for the running
coupling and $\gamma\sim 0.61$ for the fixed coupling solution. The
differences between fixed and running coupling solutions at $\tau>1$
are evident from \fig{scal}.  This is a puzzling result that remains
to be understood from purely analytical methods.

%%%%%%%%%%%%%%%%%%%%%%%%%%%%%%%%%%%%%%%%%%%%%%%%%%%%%%%%%%%%%%%

\subsection{Subtraction Term}
\label{subterm}

%%%%%%%%%%%%%%%%%%%%%%%%%%%%%
\begin{figure}[t]
\begin{center}
\includegraphics[height=10cm]{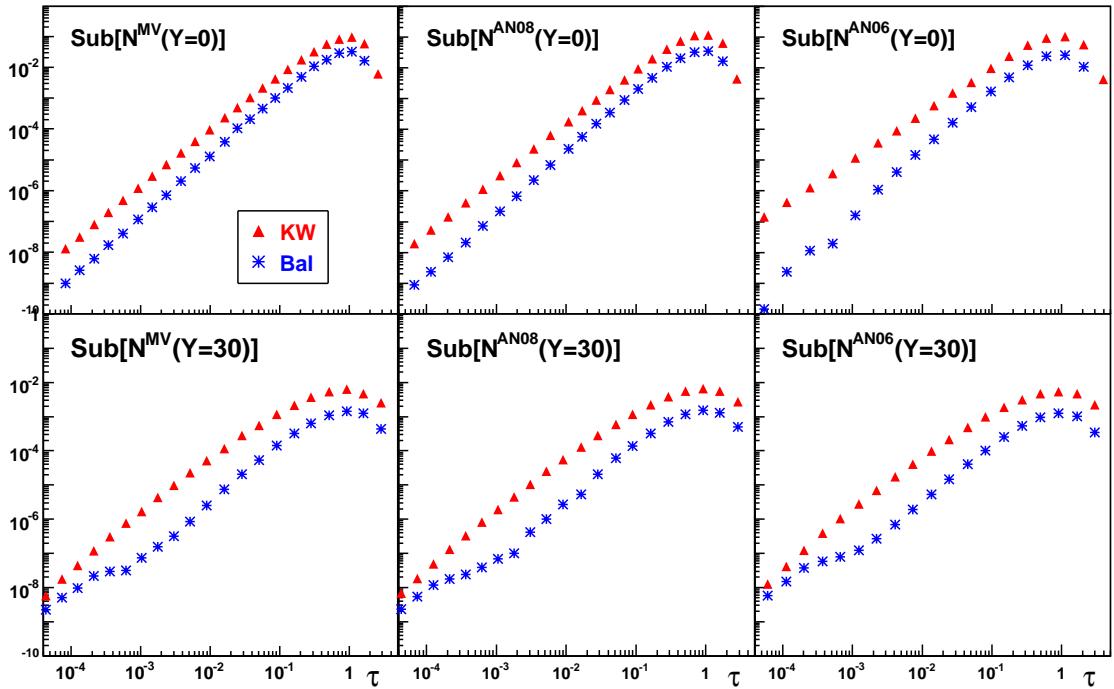}
\caption{Subtraction contribution calculated in the KW scheme
  (triangles) and in Balitsky's (stars). The trial functions
  correspond to the solutions of the evolution under Balitsky running
  coupling scheme at   rapidities $Y=0,30$ for MV (left), AN08
  (center) and AN06 initial conditions.} 
\label{sub1}
\end{center}
\end{figure}
%%%%%%%%%%%%%%%%%%%%%%%%%%%%%%

Before attempting to solve the complete evolution equation, and in
order to gain insight in the nature and structure of the subtraction
contribution, we first evaluate the subtraction functional for both
Balitsky, \eq{sub_full_bal}, and KW, \eq{sub_full_kw}, schemes using a
set of trial functions for $S$ which we choose to consist of the
solutions of the evolution equation with the running coupling in
Balitsky's scheme at different rapidities and of the three initial
conditions considered above in this work.

%%%%%%%%%%%%%%%%%%%%%%%%%%%%%
\begin{figure}
\begin{center}
\includegraphics[height=11cm]{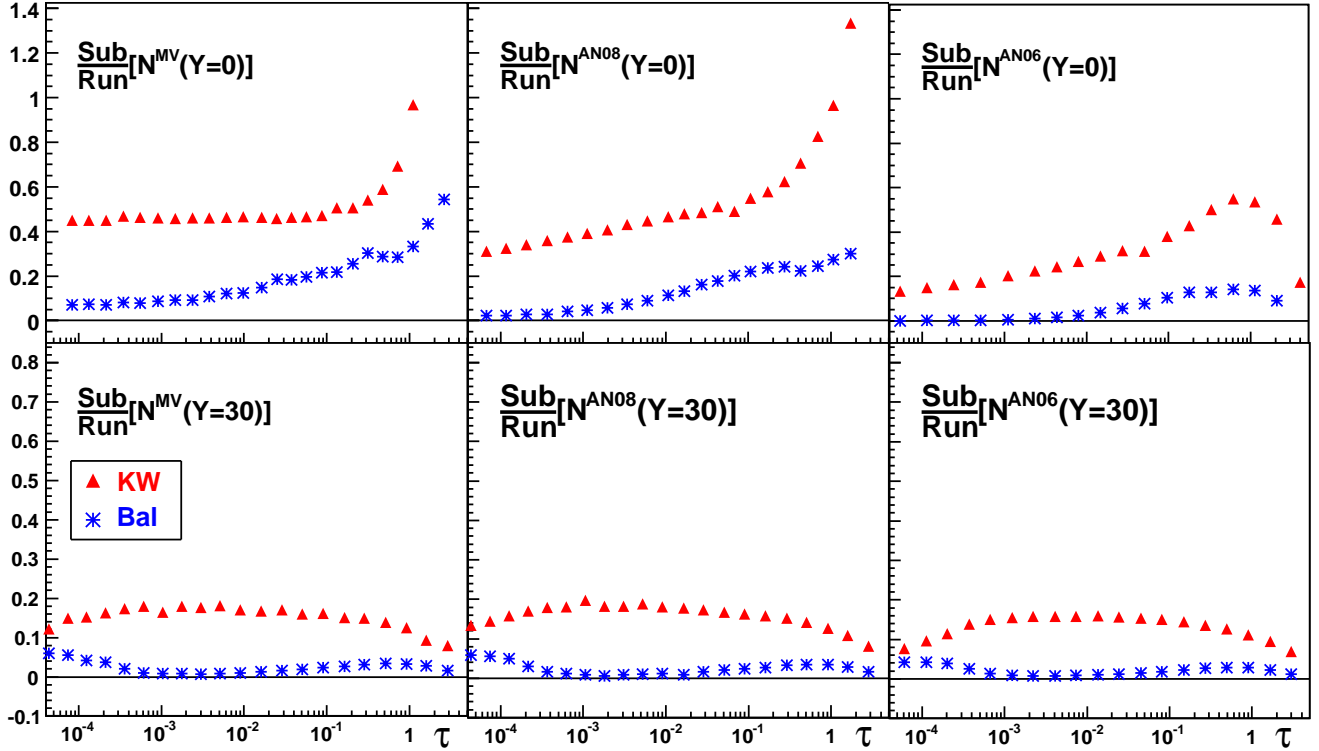}
\end{center}
\caption{Ratio of the subtraction over the running terms,
  $\mathcal{D}(r,Y)=\mathcal{S}[N(r,Y)]/\mathcal{R}[N(r,Y)]$,
  calculated in both KW (triangles) and Balitsky (stars) schemes for
  MV (left), AN08 (middle) and AN06 (right) initial conditions at
  rapidities Y=0 (top) and Y=30 (bottom).}
\label{subr}
\end{figure}
%%%%%%%%%%%%%%%%%%%%%%%%%%%%%%

Two main remarks can be made about our results, shown in \fig{sub1}:

i) For all the trial functions considered in this work, the
subtraction contribution is much larger in the KW scheme than in
Balitsky's.  A plausible explanation for this is that Balitsky's
subtraction contribution, \eq{sub_full_bal}, when expanded in terms of
dipole scattering amplitudes, $N=1-S$, reduces to a sum of non-linear
terms, since all the linear terms in the expansion cancel each other
due to the ${\un z}_1 \leftrightarrow {\un z}_2$ symmetry of the
kernel, whereas in the KW case no such cancellation happens and the
subtraction contribution, \eq{sub_full_kw}, also includes linear
terms, which are dominant over the non-linear ones in the
non-saturated domain where $N \ll 1$.

ii) The subtraction contribution $\mathcal S$ has the same sign as the
running coupling contribution $\mathcal R$ in the whole $\tau$ range
which, together with the relative minus sign assigned to the
subtraction term in \eq{frs}, implies that the proper inclusion of the
subtraction term reduces the value of the functional that governs the
evolution, $\mathcal{F}$. In other words: the subtraction contribution
tends to systematically slow down the evolution, as we shall
explicitly confirm in the next subsection.

To better quantify the size of the subtraction contribution, we plot
the ratio
$\mathcal{D}(r,Y)\equiv\mathcal{S}[N(r,Y)]/\mathcal{R}[N(r,Y)]$ in
\fig{subr}. At $Y=0$, the relative weight of the subtraction
contribution with respect to the running one within the KW scheme and
for a MV initial condition goes from a $\mathcal{D}\sim 0.4$ at small
$\tau$ to $\mathcal{D}\sim 1$ at $\tau\sim 1$. The same ratio for the
Balitsky scheme takes significantly smaller values: it goes from
$\mathcal{D}\sim 0.1$ at small $\tau$ to $\mathcal{D}\sim 0.4$ for
$\tau\sim1$.  As the evolved solutions get closer to the scaling
function, i.e. for larger rapidities, the $r$ dependence of the ratio
becomes flatter and its overall normalization goes down to an
approximately constant value $\mathcal{D}\sim 0.15$ for the KW scheme
and $\mathcal{D}\sim 0.025$ for that of Balitsky. This behavior
remains unaltered when going from rapidity $Y=20$ to $Y=30$, which
suggests that the ratio may saturate to a fixed value in the
asymptotic region.

%%%%%%%%%%%%%%%%%%%%%%%%%%%%%
\begin{figure}[bht]
\begin{center}
\includegraphics[height=10cm]{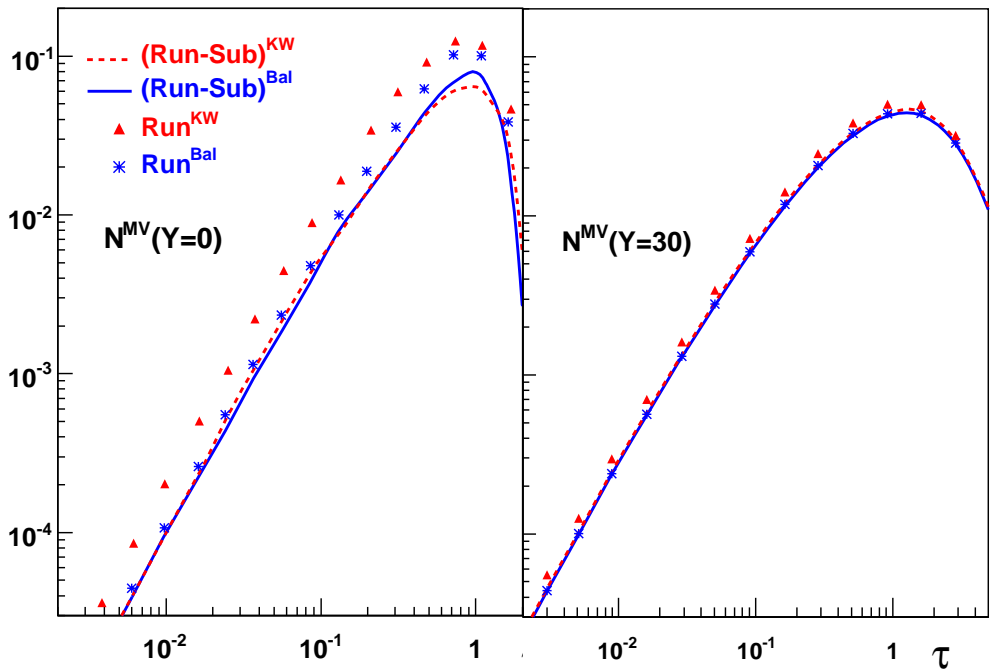}
\end{center}
\caption{Total kernel $\mathcal{F}=\mathcal{R}-\mathcal{S}$ calculated
  under Balitsky's scheme, Eqs (\ref{kbal}) and (\ref{sub_full_bal}),
  (solid line) and under the KW scheme, Eqs (\ref{kkw}) and
  (\ref{sub_full_kw}), (dashed line). The overlap of the two lines
  shows the agreement between the two calculations. Triangles stand
  for the running coupling term calculated in the KW approach,
  $\mathcal{R}^{\text{KW}}$, while stars stand for the running
  coupling term under Balitsky's scheme, $\mathcal{R}^{\text{Bal}}$.
  The trial functions $N(r,Y)$ correspond to the solution of the
  evolution with only running coupling under Balitsky's scheme at Y=0
  (left) and Y=30 (right) for a MV initial condition.}
\label{subt}
\end{figure}
%%%%%%%%%%%%%%%%%%%%%%%%%%%%%%
Finally, we have checked that combining the subtraction and running
coupling contributions for both schemes adds up to the same result.
This is shown in \fig{subt}, where we plot the value of the total
functional $\mathcal{F}=\mathcal{R}-\mathcal{S}$ calculated under the
KW scheme (Eqs. (\ref{kkw}) and (\ref{kw_run}) for the running
coupling term, $\mathcal{R}$, and \eq{sub_full_kw} for the subtraction
term, $\mathcal{S}$) and under Balitsky's scheme (Eqs.  (\ref{kbal})
and (\ref{bal_run}) for the running coupling term and
\eq{sub_full_bal} for the subtraction term). The two results coincide
within the estimation of the numerical accuracy previously discussed.
The agreement between the two results is better in the small-$\tau$
region, where the two curves lay almost on top of each other. In the
saturation region, $\tau\gtrsim 1$, the agreement is slightly worse,
although the differences between the values of $\mathcal{F}$
calculated in both schemes is still much less than the differences
between the running coupling terms themselves. This slight remaining
disagreement between the Balitsky's and KW prescriptions may also be
due to inaccuracies in a Fourier transform of a geometric series
performed in arriving at \eq{K1run_final}.  This result serves as a
cross-check of our numerical method and as an additional confirmation
of the agreement of the independent calculations derived in
\cite{Balitsky:2006wa,Kovchegov:2006vj}.

%%%%%%%%%%%%%%%%%%%%%%%%%%%%%%%%%%%%%%%%%%%%%%%%%%%%%%%%%%%%%%%%%%%%%%%%%

\subsection{Complete running coupling BK equation}
\label{comprc}

%%%%%%%%%%%%%%%%%%%%%%%%%%%%%%
\begin{figure}[t]
\begin{center}
\includegraphics[height=9cm]{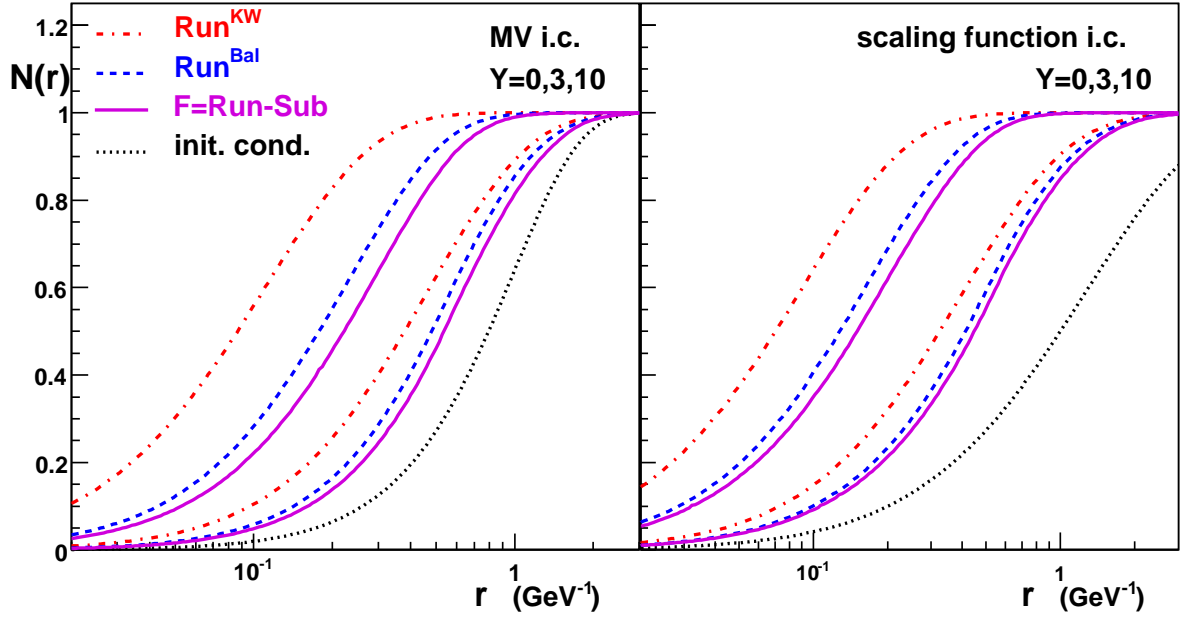}
\caption{Solutions of the complete (all orders in $\as \, \beta_2$) 
  evolution equation given in \eq{frs} (solid lines), and of the
  equation with Balitsky's (dashed lines) and KW's (dashed-dotted)
  running coupling schemes at rapidities $Y=0$, $5$ and $10$.  Left
  plot uses MV initial condition. The right plot employs the initial
  condition given by the dipole amplitude at rapidity $Y=35$ evolved
  using Balitsky's running coupling scheme and with $r$-dependence
  rescaled down such that $Q_s = Q'_s = 1$~GeV.}
\label{subev}
\end{center}
\end{figure}
%%%%%%%%%%%%%%%%%%%%%%%%%%%%%%%%%%%%%%%%%%%%%%%%%%%%%%%%%%%%

In this section we calculate the solutions of the complete evolution
equation, \eq{frs}, including both the running and subtraction terms
obtained by the all-orders $\as \, N_f$ resummation and by the $N_f
\rightarrow - 6 \pi \beta_2$ replacement.  Since the numerical
evaluation of the subtraction contribution at each point of the grid
and each step of the evolution would require an exceedingly large
amount of CPU time consumption, the strategy followed to include it in
the evolution equation consists of calculating such contribution only
in a small set of grid points at each step of the evolution, which we
fixed at $n=16$, between the points $r_{1}$ and $r_{2}$, which are
determined at each step of the evolution by the conditions
$N(Y,r_{1})=10^{-9}$, and $N(Y,r_{2})=0.99$, and then using power-law
interpolation and extrapolation to the other points of the grid.  Both
the running and subtracted terms are calculated within Balitsky
scheme.  This procedure is motivated by the fact that, as discussed in
the previous section, the subtraction contribution can be regarded as
a small perturbation with respect to the running coupling term within
Balitsky's scheme and by the fact that it is a rather smooth function
that can be well fitted by a power-law function in most of the
$r$-range.  The accuracy of this procedure has been checked by
doubling the number of points at which the subtraction contribution is
calculated at each step of the evolution, i.e. by setting n=32.  At
Y=2, the differences between the solutions obtained with the two above
mentioned choices for n were less than a $8 \%$ in the tail of the
solution, $r<r_1$, and less than a $3\%$ for $r>r_1$.
%%%%%%%%%%%%%%%%%%%%%%%%%%%%%%
\begin{figure}[t]
\begin{center}
\includegraphics[height=10cm]{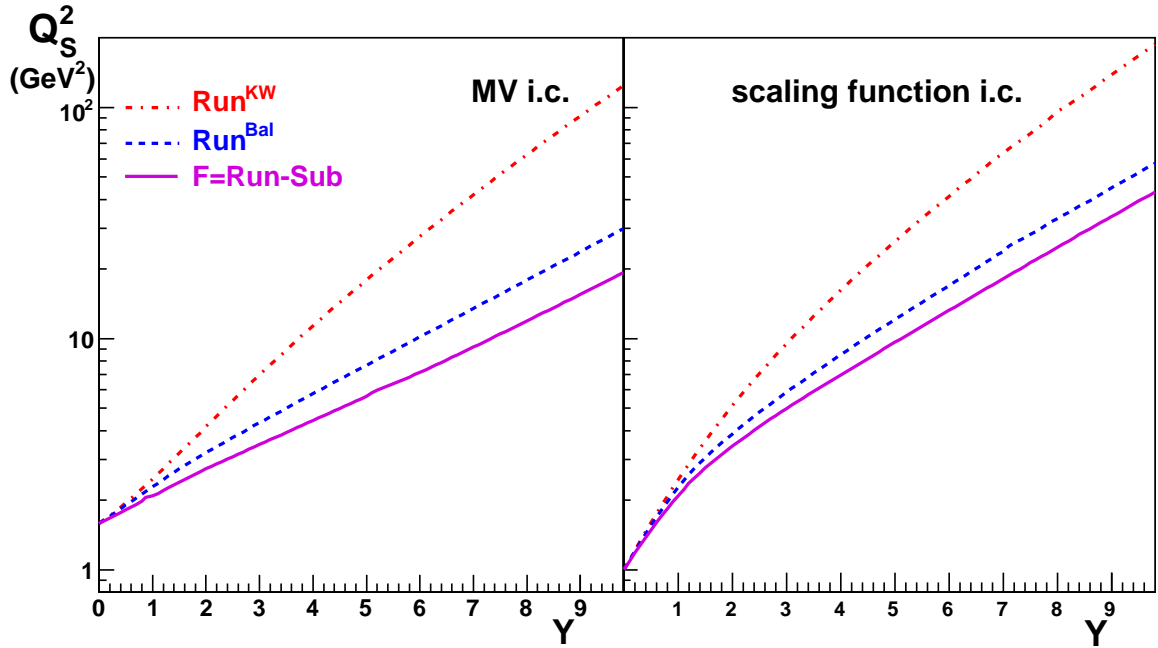}
\caption{Saturation scale corresponding to the solutions plotted in
  \fig{subev}.} 
\label{subqs}
\end{center}
\end{figure}
%%%%%%%%%%%%%%%%%%%%%%%%%%%%%%%%%%%%%%%%%%%%%%%%%%%%%%%%%%%%

The results of the evolution calculated in this way and using MV and
rescaled asymptotic running coupling solution (Y=35) as initial
conditions are plotted in \fig{subev}.  They confirm the expectations
raised in the previous Subsection: the inclusion of the subtraction
terms considerably slows down the evolution with respect to the sole
consideration of the running coupling contributions. Moreover, the
reduction in the speed of the wave front is much larger for the KW
scheme than for that of Balitsky one for both initial conditions.
However, the closer the initial condition is to the asymptotic running
coupling scaling function, the smaller are the effects of the
subtraction contribution.  These features can be better quantified by
inspecting the rapidity dependence of the saturation scale generated
by the evolution, plotted in \fig{subqs}. At rapidity $Y=10$ the ratio
of the saturation scale $Q_s$ yielded by the KW scheme to $Q_s$ given
by the complete $\as \, \beta_2$-evolution equation is a factor of
$\sim 2.5$ for the MV initial condition and a factor of $\sim 2.1$ for
the asymptotic running coupling initial condition. At the same
rapidity, the ratio of the saturation scale obtained under Balitsky's
scheme to $Q_s$ corresponding to the complete $\as \,
\beta_2$-evolution is $\sim 1.25$ for the MV initial condition and
$\sim 1.15$ for the scaling function initial condition.  Thus, in spite
of the smallness of the ratio of the subtraction terms to the running
coupling contributions at high rapidity, which is $\sim 0.025$ for
Balitsky's and $\sim 0.15$ for KW scheme at $Y=30$ (see bottom plots in
\fig{subr}), the proper inclusion of the subtraction term results in
fairly sizable effects in the solutions of the evolution equation.

%%%%%%%%%%%%%%%%%%%%%%%%%%%%%%
\begin{figure}
\begin{center}
\includegraphics[height=9cm]{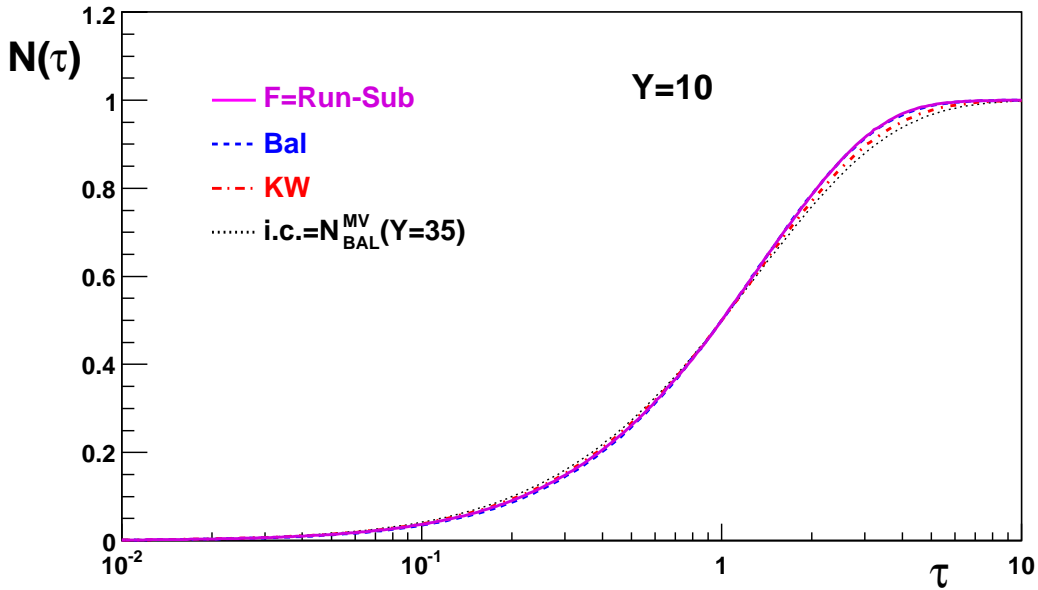}
\caption{Rescaled solutions given by the complete $\as \, \beta_2$-evolution
  equation (solid line) and for KW (dashed-dotted line) and Balitsky's
  (dashed line) running coupling schemes at $Y=10$. The initial
  condition corresponds to the dipole amplitude at rapidity $Y=35$
  evolved using Balitsky's running coupling scheme and with
  $r$-dependence rescaled down such that $Q_s = Q'_s = 1$~GeV.}
\label{scalsub}
\end{center}
\end{figure}
%%%%%%%%%%%%%%%%%%%%%%%%%%%%%%%%%%%%%%%%%%%%%%%%%%%%%%%%%%%%

Finally, we notice that the scaling behavior of the solution is not
affected by the subtraction term. This is seen in figure
\fig{scalsub}, where we evolve starting from an initial condition
already close to the running coupling scaling function and plot the
solutions of the evolution equation obtained with just running
coupling terms (see Section \ref{running}) and the solution of the
complete $\as \, \beta_2$-evolution at rapidity $Y=10$. It is clear
that, within the numerical accuracy, no departure from the scaling
behavior is observed.  Therefore the main effect of a proper
consideration of the subtraction term is the one of reducing the speed
of the evolution.  It does not violate or modify the geometric scaling
property of the solutions established in Section \ref{scaling}. In our
understanding geometric scaling appears to persist when the running
coupling effects are included because, at high enough rapidity $Q_s (Y)
\gg \Lambda$, such that the new (from the LO standpoint) momentum
scale $\Lambda$ introduced by the running coupling can be safely
neglected. Hence the dynamics is again characterized by a single
momentum scale $Q_s (Y)$. At the same time running coupling does
modify the evolution kernel, leading to a different shape of the
scaling function.

%%%%%%%%%%%%%%%%%%%%%%%%%%%%%%%%%%%%%%%%%%%%%%%%%%%%%%%%%%%%%%%%%%%%%%%%

\section{Conclusions} \label{conclusions}

In this paper we have taken into account all corrections to the
kernels of the non-linear JIMWLK and BK evolution equations containing
powers of $\as \, N_f$. We reiterated the fact that the separation of
the resulting kernel resumming all powers of $\as \, N_f$ into the
running coupling and subtraction parts, as done in the previous
calculations of \cite{Balitsky:2006wa,Kovchegov:2006vj}, is not
justified parametrically. We have then performed numerical analysis
with the following conclusions.

\begin{itemize}
\item First we solved the evolution equations derived in
  \cite{Balitsky:2006wa} and \cite{Kovchegov:2006vj} keeping only the
  running coupling part or the evolution kernel and neglecting the
  subtraction term. Comparing to the results for fixed coupling
  obtained in \cite{Albacete:2004gw} we confirmed the conclusion
  reached in \cite{Albacete:2004gw} that the growth with rapidity is
  substantially reduced when running coupling corrections are
  included.  The results for three different initial conditions are
  shown in \fig{sols}. We observe that the solution of the equation
  derived in \cite{Kovchegov:2006vj} differs significantly from that
  derived in \cite{Balitsky:2006wa}, but agrees (with good numerical
  accuracy) with the solution of the BK evolution equation with the
  coupling running at the parent dipole size. (The latter is just a
  model of the running coupling not resulting from any calculations,
  which we plot for illustrative purposes.) We also observe that at
  sufficiently high rapidity both equations from
  \cite{Balitsky:2006wa} and from \cite{Kovchegov:2006vj} give us the
  same scaling function for the dipole amplitude $N (r, Y)$ as a
  function of $r \, Q_s (Y)$, which is also in agreement with the
  scaling function given by the parent dipole running, as shown in
  \fig{scal}. The fact that the scaling is preserved when the running
  coupling corrections are included was previously established in
  \cite{Albacete:2004gw}, though for models of running coupling only.
  The shape of the scaling function is very different from that
  obtained from the fixed coupling evolution equations. In particular,
  we found that for dipole sizes below $0.1 / Q_s$ the anomalous
  dimension of the scaling function in the running coupling case
  becomes $\gamma \approx 0.85$ (see \fig{anom}). This is different
  from the result of several analytical estimates
  \cite{Mueller:2002zm,Triantafyllopoulos:2002nz,Iancu:2002tr,Munier:2004xu,Beuf:2007cw},
  which expect the anomalous dimension not to change when running
  coupling corrections are included and to remain at its fixed
  coupling value of $\gamma \approx 0.63$.
  
\item We have then evaluated the subtraction term for both
  calculations performed in \cite{Balitsky:2006wa} and
  \cite{Kovchegov:2006vj}.  We demonstrated that subtracting the
  subtraction terms from the running coupling terms makes the full
  answer agree for both calculations of \cite{Balitsky:2006wa} and
  \cite{Kovchegov:2006vj}, as shown in \fig{subt} for the right hand
  side of the evolution equation.  It turns out that the subtraction
  term ${\mathcal S}^{\text{Bal}} [S]$, which has to be subtracted
  from the result of \cite{Balitsky:2006wa}, is systematically smaller
  than ${\mathcal S}^{\text{KW}} [S]$, to be subtracted from the
  result of \cite{Kovchegov:2006vj}, over the whole rapidity range
  studied here.  This implies that the result of
  \cite{Balitsky:2006wa} should have a smaller correction than the
  result of \cite{Kovchegov:2006vj} and is thus closer to the full
  answer. The subtraction terms ${\mathcal S}^{\text{Bal}} [S]$ and
  ${\mathcal S}^{\text{KW}} [S]$ are plotted in \fig{sub1} as
  functions of the dipole size $r$ for different values of rapidity.
  Their relative contributions to the evolution kernel are shown in
  \fig{subr}, where we plotted the subtraction functional divided by
  the running coupling functional. From those figures we conclude that
  both the magnitude of these extra terms and their relative
  contribution to the evolution kernel decrease with increasing
  rapidity. Hence, while at "moderate" rapidities (the ones closer to
  realistic experimental values) the subtraction term is important for
  both calculations \cite{Balitsky:2006wa,Kovchegov:2006vj}, it
  becomes increasingly less important at asymptotically large
  rapidities. The physics is easy to understand: the subtraction terms
  are $o(\as^2)$, while the running coupling part of the kernel is
  $o(\as)$. Hence, if we suppose that the effective value of the
  coupling is given by its magnitude at the saturation scale $Q_s
  (Y)$, then, as rapidity increases, the coupling would decrease,
  making the subtraction term much smaller than the running coupling
  term. Indeed, while at asymptotically high rapidities the assumption
  of \cite{Balitsky:2006wa,Kovchegov:2006vj} that the subtraction term
  could be neglected is justified, making the results of
  \cite{Balitsky:2006wa} and \cite{Kovchegov:2006vj} agree with each
  other, for rapidities relevant to modern days experiments the
  subtraction term is numerically important.
 
\item With the last conclusion in mind we continued by numerically
  solving the full evolution equation resumming all powers of $\as \,
  N_f$ in the evolution kernel, which now would combine both the
  running coupling and the subtraction terms. The five-dimensional
  integral in the subtraction term (\ref{sub_full}) made obtaining
  this solution rather difficult. The outcome of the calculation is
  shown in \fig{subev}. All the main conclusions stated above were
  again confirmed by the solution of the full equation. At
  asymptotically high rapidity scaling regime is recovered, as can be
  seen from \fig{scalsub}. As the subtraction term is less important
  in that regime, the scaling function appears to be the same as in
  the case of having only the running coupling term in the kernel. The
  anomalous dimension again turns out to be $\gamma \approx 0.85$, in
  disagreement with the analytical expectations of
  \cite{Mueller:2002zm,Triantafyllopoulos:2002nz,Iancu:2002tr,Munier:2004xu,Beuf:2007cw}.
  However, the scaling of the saturation scale with rapidity appears
  to be in agreement with the expectations of analytical work of
  \cite{Mueller:2002zm,Triantafyllopoulos:2002nz,Beuf:2007cw}, as
  shown in \fig{subqs}.
\end{itemize}

We conclude by observing that the knowledge of the non-linear
small-$x$ evolution equation with all the running coupling corrections
included brings us to an unprecedented level of precision allowing for
a much more detailed comparison with experiments than was ever possible
before.

%%%%%%%%%%%%%%%%%%%%%%%%%%%%%%%%%%%%%%%%%%%%%%%%%%%%%%%%%%%%%%%%%%%%%%%%%%%%%%%

\section*{Acknowledgments} 

We would like to thank Heribert Weigert for many informative and
helpful discussions at the beginning of this work. A portion of the
performed work was motivated by stimulating discussions with Robi
Peschanski, which we gratefully acknowledge.

This research is sponsored in part by the U.S. Department of Energy
under Grant No. DE-FG02-05ER41377. This work was supported in part by
an allocation of computing time from the Ohio Supercomputer Center.

%%%%%%%%%%%%%%%%%%%%%%%%%%%%%%%%%%%%%%%%%%%%%%%%%%%%%%%%%%%%%%%%%%%%%%%%%%%%%%

%\bibliography{references}                   
%\bibliographystyle{JHEP}
\providecommand{\href}[2]{#2}\begingroup\raggedright\endgroup
\end{document}